# An ab-initio derivation to discuss the heterodyne versus direct detection decision problem for astronomical infrared interferometry

E. A. Michael[1,2,*], F. E. Besser[3,2], M. Hadjara[4,2], E. Moreno[5,2], A. Berdja[6,2], M. Piña[2], G. Pereira[2]

[1] 1. Physics Institute, Universität zu Köln, Cologne, Germany, email: michael@ph1.uni-koeln.de
[2] Astro-Photonics Laboratory, Dept. of Electrical Engineering, FCFM, Universidad de Chile, Santiago, Chile
[3] Las Campanas Observatory, Raúl Bitrán 1200, La Serena, Chile
[4] Chinese Academy of Sciences South America Center for Astronomy, National Astronomical Observatories, CAS, Beijing 100101, China
[5] Departamento de Física y Matemáticas, Facultad de Ciencias, Campus Científico-Tecnológico de la Universidad de Alcalá, Calle el Escorial, 19-21, 28805 Alcalá de Henares, Madrid, Spain
[6] DICTUC S.A., Pontificia Universidad Católica, Santiago, Chile

## ABSTRACT

A consistent and explicit spectral comparison between heterodyne (HD) and direct detection (DD) derived from first principles including the atmospheric transmission and low beam-filling factors could not be found yet in literature but is needed for decisions in technology planification for future infrared interferometry facilities which are e.g. focused on planet formation. This task requires both, high sensitivity continuum and Doppler-resolved emission and absorption line detection in the mid-IR range (N- and Q-bands) at lower source temperatures (300-1000 K).

The signal-to-noise ratios (SNRs) are derived for both schemes within the same semi-classical theory, which consists of classical mode theory for coupling to an antenna and occupation of these modes by quanta of three radiation fields, the thermal signal, the thermal background, and for HD also the coherent local oscillator (LO). The effects of very small beam filling factors (interferometry) and atmospheric absorption/emission could be consistently incorporated this way, as well as quantum-noise propagation which allows in HD the consideration of balanced mixers with cross-correlation (CC). Especially, the transition from pre- to post-detection SNRs was considered meticulously. We do this all because the usually cited SNR-expressions were derived for a too simple and unrealistic case, and moreover contain some wrong assumptions. The actual expressions for both detection schemes are plotted together over wavelength and source power for selected cases.

For interferometric observations of warm sources very small against the single-telescope beam (very small beam filling factors), heterodyne can already be better than DD for wavelengths longer than 5 microns even with a smaller bandwidth, because the signal-to-noise ratio (SNR) in DD is more affected by the ambient temperature background than it is in HD: Firstly, for weak signal powers compared to a local thermal background, the power law for the SNR for DD turns over to quadratic, whereas for HD it stays linear. Secondly, in DD interferometry the unavoidable distribution onto many pixels reduces additionally the SNR for signals which are weak against dark and read-out noise, which introduces a SNR-penalty for DD at higher resolutions. Thirdly, in CC of HD using balanced mixers (balanced correlation receivers, BCR), we consider a recently discovered SNR-improvement factor of 10-20 in comparison to auto-correlation (AC) of a single-telescope HD receiver. For λ=10-20 μm (N- and Q-bands), considering a single baseline, a HD BCR interferometry system should already extend the limiting sensitivity by 3-4.5 astronomical magnitudes compared to a high-resolution DD interferometry system with the same channel-width, and for more baselines this difference should increase.

As a result, we can introduce a novel HD scheme for astronomical interferometry gaining an order of magnitude in sensitivity against conventional HD and calculate that it should trespass the sensitivity of DD interferometry in the N- and Q-bands for a spectral resolution of R=10000, and should do also for R=300 with doable technical improvements. In view of the many advantages of heterodyne interferometry at large baselines and telescope numbers, this result encourages to develop broad-band heterodyne technologies for future mid-infrared interferometry facilities and for new instruments at existing facilities.

## 1. Introduction

Astronomy aims to detect ever weaker signals with affordable integration times and telescope sizes. Quantum statistics [1] and the vacuum fluctuations of the electromagnetic field [2] establish fundamental limitations to sensitivity. Here, incoherent and coherent detection are competing to achieve the higher signal-to-noise ratio (SNR), depending on wavelength and other parameters.

### 1.1. Two competing detection principles

In direct (incoherent) detection the signal photons alone directly generate photoelectrons. The sensitivity is then limited only by the Poisson variance of the signal photons integrated on the detector pixel within a time and frequency interval [1], but substantial post-detection amplifier noise adds to this. High spectral resolution ($> R = 10^5$) is achievable only with bulky (collimated beam) wavelength-selective optics in front of the detectors [3][4][5], which implies increasing losses towards higher spectral resolution.

In heterodyne (coherent) detection, the weak electromagnetic field to be detected is mixed on a fast detector (the mixer) with a strong monochromatic reference signal, the "local oscillator" (LO), down-converting the sidebands into the intermediate frequency (IF) band at $\nu_{IF} =$ $|\nu_S - \nu_{LO}|$, preserving their phases. Therefore, in principle, the signal can be amplified in the very moment of detection so highly by multiplication with the strong LO that the impact of post-detection amplifier noise is eliminated, although in practice limitations are given for the LO-power and by mixing losses. Unfortunately, this brings in fundamental quantum noise from zero point fluctuations (ZPF), resulting in $h\nu/2$ of white noise power per Hertz and mode [2][6], equivalent to the emission of a thermal source of "noise temperature" $T = h\nu/2k_B$. ZPF cannot be detected by an incoherent detector [7], but they add inside a coherent receiver and any phase preserving amplifier to the source noise seen by it [8]. In case of a local oscillator (LO), ZPF have the same effect on the final SNR as adding the shot noise to the LO, which is the approach in this paper.

In DD, photons enter both telescopes simultaneously and interfere with themselves in the beam-combination instrument, by being locally detected at a single pixel of the detector-array. This way, many are needed to form the interference pattern, to make possible the measurement of its contrast, the visibility value ($V$), and the relative phase between the two partial waves [9]. In HD, the photons are first converted to photoelectrons locally at the antennas, and so the interference is measured over the cross-correlation (CC) coefficient (equivalent to visibility) of the electron-fluctuation amplitudes in the mixer. Therefore, both detection principles need a similar number of photons to measure the interference with a similar signal-to-noise ratio, surprisingly despite of the limit in HD by the ZPF.

**1.2. Combined high spatial and spectral resolution**

When the highest angular and/or spectral resolution is necessary, the difficulties of interferometry and/or high-resolution spectroscopy add to the difficulty of high sensitivity. Interferometry has become a key technology in optical [10] and submillimeter [11] high spatial resolution astronomy. In optical and infrared heterodyne interferometry, the highest possible visibility-sensitivities are required at the largest possible baselines to surpass the spatial resolution so far achieved in the submm-range, which calls for breaking records in detection sensitivity.

The main reasons for using HD in mid-infrared interferometry would be: 1.) to carry over the spectral high-resolution analyzing power of submm-interferometers (e.g. ALMA, VLA and IRAM) for chemical characterization; 2.) to resolve low-velocity structures in accretion disks analyzing spectral line shifts (line profile information) and therefore to be able to resolve Hills-spheres around forming planets, at typical velocities smaller than that of Earth (<30 km/s, R>10000), and also under near-to face-on orientations of the accretion disks (R>100000 – then only doable by heterodyne); 3.) to realize kilometric baselines similar to those of ALMA, fibers for the distribution of the local oscillator laser have to be used. To realize in future even larger baselines (10-100 km) without fiber connections for the LO, it should be considered that sufficiently precise atomic clocks will be available to synchronize the local oscillators at the individual telescopes to perform very-long baseline interferometry even in the infrared range. These so-called optical clocks already have reached ×100 higher precision (lower relative Allan deviation) than the best Rubidium atomic clock [12]. Considering that optical clock fiber networks already have been demonstrated over many hundreds of kilometers [13], heterodyne technology should be able soon to allow for larger than few-kilometer baselines in the infrared, similar to what was demonstrated recently over global distances in the sub-millimeter range with the use of hydrogen maser atomic clocks to image event horizons of black holes [14]. Space-born time-delay heterodyne interferometry is already being conceived for baselines of millions of kilometers [15]. 4.) One of the main advantages of HD interferometry is the ability to "clone" the IF-signal without any punishment in SNR-loss, and so the realization of large arrays is possible (see e.g. ALMA).

To the contrary, in DD interferometry the division of the optical power at the beam-combiner instrument leads to a sensitivity down-cut increasing towards larger arrays. This is not discussed in the current paper which focuses solely on the comparison of sensitivities for single telescopes and two-telescope interference. Furthermore, the beam transmission from two or more telescopes to the interference point cannot bridge more than a few kilometers due to beam divergence and reimaging, and a significantly reduced throughput due to a high number of mirrors in beam relaying cannot truly be avoided. In any case, the "étendue" (optical throughput = aperture × opening angle) must be restricted to the fundamental mode to not blur the interference pattern. Additionally, an array of detectors is needed to image the fringes in the superposition of the beams from two or more telescopes. (Alternatively, a single-pixel detector could be used with several exposure times with different phases between the telescopes.) Therefore, the signal power of a point-like source (small in angular extension against the single-telescope point-spread function, the "PSF") needs to be diluted over multiple detectors, which leads to a SNR-loss in case of readout noise, see analysis in the appendix 8.4. For example, the MATISSE-instrument uses 72 pixels for the detection of the fringe pattern combining 4 telescopes [16]. Some camera arrays with suppressed readout noise have been developed so far, e.g. EMCCDs in the visible and e-APDs around 3 µm, but in the mid-infrared (8 – 30 µm) this is not yet within reach.

On the other hand, in heterodyne interferometry all signal power is concentrated on only one detector (two in balanced, and four in balanced sideband separation) per telescope. Moreover, we have demonstrated in the laboratory that by using as the detection signal the correlation of two balanced heterodyne receivers, we can reach the quantum limit for the noise temperature, and possibly can trespass it even by a factor of 3-4 [17], [18]. Fast back-end electronics in correlation becomes steadily easier to realize, also here "Moore's law" applies: Nowadays, 2.5 GHz bandwidth correlation is available for 15kUSD (ROACH2-board), and in a couple of years 25 GHz systems will be available for probably a similar price [19]. Maybe quantum-computing will come in later. On the above development it will depend if proposed frequency-resolving ultra-broad-band photonic correlator schemes stay attractive [20], [21]. Also interesting is the development of dispersed heterodyne receivers [22].

**1.3. The need for better comparison of HD versus DD**

Regarding a general discussion of the sensitivities of single-telescope HD versus DD, there are only very few references in the literature: Harris briefly discussed it for the better signal to noise ratio in case of single telescope sub-millimeter wave detection in astronomy [23]. Brown analyzed exhaustively both, DD and HD, in the sub-millimeter range for different cases in remote sensing, but unfortunately did not contrast both schemes quantitatively against each other for astronomy [24]. Chen et al. have given a comparison for optical inter-satellite communication links [25], while Chabata et al. [26] and Gatt et al. [27] have compared both detection schemes for LIDAR. However, none of them worked out a generalized first-principles scheme for the comparison over all wavelengths, and by far not for the case of ground-based astronomical interferometry, including atmospheric absorption/emission and small beam-filling factors. Thus, it was found that the literature has not treated yet really the decision problem between HD and DD in astronomical interferometry. The parts necessary for it are distributed under many different notations, some parts are still missing, and some clarifications and corrections now appear to be necessary. Usually, the SNR-formulas appeared as higher-order citations seemingly not derived properly anywhere from first principles.

In view of the plans for new mid-infrared interferometry instrumentation and facilities it appears therefore that such a pending full discussion is now definitely needed. It requires a more comprehensive, integrating, and careful ab-initio derivation of the theory, to have it in a form to properly compare DD and HD for astronomy as a function of wavelength. Some effects previously have not been considered, as the small beam-filling factors prevalent in interferometry, the atmospheric absorption/emission, in DD the dilution over many pixels, and in HD the suppression of background radiation and receiver-/LO-noise in CC.

Townes et al. presented simplified noise considerations for one telescope looking at an extended source, i.e. source-to-telescope beam-overlap of unity, [28], [29], [30], which are in a way based on Kingston [31] and Teich [32], but not completely, and the difference (the integration time dependence) is not derived. Atmospheric absorption/emission and source-to-telescope beam-overlap was not included. Furthermore, a square-law dependence of the HD CC-SNR on the single-telescope SNR was proposed without derivation, on which one cannot agree from simple reasons. Own recent laboratory measurements showed a linear dependency for CC like it is observed for AC [17].

Two different starting concepts were reported seemingly excluding each other, one concept including directly the integration time [28], [29], and [30], on which the variance of the white noise should depend linearly, and so the SNRs with the square-root, and another concept without it [33], [34]. To our knowledge it was not made clear in the previous literature how these two are related. Both do not start out from the radiation noise, nor consider its propagation through the detection process, but rather start from the post-detection dc-current shot noise. From the theory presented here we can now understand the latter of both concepts as the post-detection single temporal mode SNR expression.

Therefore, we wondered how to start off the derivations for both, HD and DD, in the way Blaney [33] did just for HD alone and Spears [34] proposed for HD and DD in parallel but without writing it out. The idea was to start with all radiation noise components, propagate them through the detection process, and add post-detection noise. Single temporal radiation modes are considered as the natural source of radiation noise, which also leads to a straightforward derivation of the square-root integration time and radiation bandwidth dependence. This developed into the present comprehensive review in unified notation which puts us into the position to critically discuss previously published and frequently cited SNR-expressions. We do this not only for single-mixer or -pixel receivers/detectors (single telescopes), but also for balanced heterodyne receivers and many-pixel direct detectors, furthermore for two-telescope cross-correlation/interference.

This procedure is structured as follows: In chapter 2 we summarize the needed quantities and relationships of radiation and its noise, and in chapter 3 we apply this to the post-detection SNRs. To keep this shortest possible, lengthy side-derivations were moved into the appendix. A discussion summarizes nine achievements of this paper.

## 2. Fundamental Relations

### 2.1. Thermal radiation signals

The total number of modes a detector interacts with is given from cavity mode wave theory [1] by the mode number per volume $\Delta V$, solid angle $\Delta\Omega$ and spectral interval $\Delta\nu$, $m_{cav,ang} = 2\nu^2/c^3$ (for two polarizations) as:

$$\Delta M_{cav,ang} = 2\left(\frac{\nu^2}{c^3}\right) \Delta V \, \Delta\Omega \, \Delta\nu = 2\left(\frac{1}{\lambda}\right)^2 \Delta A \, \Delta\Omega \, \Delta\nu \left(\frac{\Delta L}{c}\right)$$
$$=: 2 \, \Delta M_A \, \Delta M_L \quad (1a)$$

where $\Delta\nu$ is the optical bandwidth and $\lambda$ is the wavelength.

$$\Delta M_A = \frac{\Delta A \Delta\Omega}{\lambda^2} \geq 1 \quad (1b)$$

is the number of transverse modes the detector or antenna "sees" (étendue, $\underline{A}$ntenna theorem), with $\Delta A$ the effective area of the beam on the detector or antenna and $\Delta\Omega$ the opening angle of the beam.

$$\Delta M_L := \Delta\nu \Delta t \quad (1c)$$

is the number of $\underline{L}$ongitudinal traveling wave modes arriving at the detector (from one direction) in the integration time $\Delta t = \Delta L/c$, so that the considered mode volume $\Delta V$ depends on the integration time. The latter are also referred to as "temporal modes" [35] [36], whereas the former as "spatial modes". A single temporal mode, $\Delta M_L = 1$, is measured if the integration time is (theoretically) as short as the coherence time, $\Delta t = t_{coh} := 1/\Delta\nu$.

The power fluctuation bandwidth $\Delta f$ is related to the integration time like

$$\Delta f = 1/2\Delta t \quad (1d)$$

Therefore, for a single temporal mode the corresponding fluctuation bandwidth is to be set to $\Delta f = \Delta\nu/2$. Each of the subsequent temporal modes has a different photon occupation number $n$ and a different phase, and so radiation noise is transported.

In case of extended sources (larger than the antenna beam) the detected source (index = S) power <u>in one polarization</u> is the energy of all the temporal modes seen during their waves are passing the antenna (HD) or detector pixel (DD) in the time $\Delta t$:

$$P_S = \varepsilon_S \Delta M_A \, h\nu \, \bar{n}_S \, \Delta\nu_S =: P_{S,\nu} \, \Delta\nu_S \qquad (2a)$$

where $\bar{n}_S$ is the "mode occupation number" which is the number of photons in the mode, specified more below. We assume from here on for simplicity that the emissivity of the astronomical sources is unity, $\varepsilon_S = 1$, but the background emissivity in an atmospheric spectral window and of the telescope and receiver optics is rather low, e.g. in the 10μm-window it might be something like $\varepsilon_B = 0.1$ [37].

In spatial interferometry very-small-extension (point-like) sky sources with respect to the single-telescope beam (or the angular telescope resolution) are observed, so that we want an optical system with "étendue" factor $\Delta M_A := 1$, i.e. a single-mode system. Therefore, in DD interferometry single-mode fibers are even inserted to ensure that from both telescopes only the fundamental modes interfere with each other, see for example the GRAVITY [38] instrument at ESO's VLTI, Paranal, Chile. Then, the étendue-advantage of DD over heterodyne is not exploitable anyway. So we have to consider a "source beam filling factor" (as known from radio astronomy), $\eta_S := \Delta\Omega_S/\Delta\Omega_T < 1$, in the sense of an overlap (photon transmission probability) to the fundamental receiver mode (single mode), where $\Delta\Omega_S = A_S/r^2$ and $\Delta\Omega_T = \lambda^2/A_T$ (with the effective Gaussian beam area of the telescope, $A_T$). The detected (polarized) power from the source is then reduced to:

$$P_S = \mathcal{T} \, \eta_S \, h\nu \, \bar{n}_S \, \Delta\nu =: P_{S,\nu} \, \Delta\nu_S \qquad (2b)$$

where $\mathcal{T}$ is the total transmission through the atmosphere and the optics from the telescope to the detector in the instrument ("throughput"), and we assume in the following for simplicity that both are at a single temperature. Effectively, the atmosphere can be approximated by $T \approx 290$ K. From radiative transfer it can be determined (see e.g. [39]) that the relative background emissivity is equal to the absorption along the optical path: $\mathcal{E}_B = \mathcal{A} = 1 - \mathcal{T}$. Therein it is $\mathcal{T} = e^{-\tau}$, where $\tau = \int_{s_0}^{S} \alpha_\nu ds$ is the optical depth in backward integration from the observer side.

On the other hand, the power received by a telescope can be expressed as

$$P_S = \mathcal{T} \, f_\nu \, A_T \, \Delta\nu \qquad (2b')$$

where $f_\nu$ is the flux, which can be expressed in terms of the astronomical magnitude $m_{AB}$ of the source as

$$f_\nu = f_{0,\nu} \cdot 10^{-\frac{m_{AB}}{2.5}} \text{ Jy} \qquad (3a)$$

$$= \eta_S \, h\nu \, \bar{n}_S/A_T \leq \frac{\Delta\Omega_S}{\lambda^2} \, h\nu \, \bar{n}_S$$

$$m_{AB} = -2.5 \log_{10}\left(h\nu \eta_S \bar{n}_S/(A_T \cdot f_{0,\nu} \cdot 10^{-26})\right) \qquad (3a')$$

and is measured in the unit Jansky (1 Jy = $10^{-26}$ W Hz$^{-1}$m$^{-2}$) [40]. In the visible it is defined $f_{0,vis} =$ 3631 Jy. $f_{0,\nu}$ drops with the wavelength, as it was defined that a typical star (e.g. Sirius) shall have a wavelength-independent magnitude regardless of its Planck-spectrum [41].

$\eta_S$ can be interpreted as the fraction $\Delta M_S$ of a hypothetical single mode which is emitted by the source with the distance $r$ and the (single-mode) area $A_S$ into the solid angle subtended by the telescope area $A_T$ as viewed from the source $(\Delta\Omega_T)$: $P_S = P_{S,\nu} \Delta\nu = \Delta M_S \, h\nu \, \bar{n}_S \, \Delta\nu$, with

$$\Delta M_S := \frac{A_S \Delta\Omega_T}{\lambda^2} = \frac{A_S}{\lambda^2}\frac{A_T}{r^2} = \frac{A_T}{\lambda^2}\frac{A_S}{r^2} = \frac{\Delta\Omega_S}{\Delta\Omega_T} = \eta_S$$

It therefore has the meaning of an overlap between two modes, a concept which is subsequently used to calculate the quantum noise propagation (see 2.3.1).

Detectors in ground-based telescopes see also a (multi-) mode-filling background radiation power from the atmosphere and the warm telescope optics:

$$P_B = P_{B,\nu} \, \Delta\nu = \mathcal{E}_B \Delta M_A \, h\nu \, \bar{n}_B \, \Delta\nu \qquad (4a)$$
$$= (1 - \mathcal{T}) \Delta M_A \, h\nu \, \bar{n}_B \, \Delta\nu$$

Even for the best case of single-mode filtering, $\Delta M_A \to 1$, this overwhelms the signal power for $\eta_S \ll 1$, when leaving the visible towards mid-infrared wavelengths [37].

$$P_{rad} = P_S + P_B \approx P_B \qquad (4b)$$

Therefore, sky position-switching on and off the astronomical source must be applied, and in case of heterodyne detection of spectral lines frequency-switching can additionally be applied.

The so-called "mode occupation number", $\bar{n}_S$ or $\bar{n}_B$, ($\overline{(...)}$ means time-averaged) the photon number per received mode (spatial and temporal), is for thermal continuum light fields of source (i=S) and background (i=B) (Planck radiation law)

$$\bar{n}_i = \left(e^{\frac{h\nu}{kT_i}} - 1\right)^{-1} \qquad (5)$$

For HD we must provide additionally a local-oscillator (LO) power to the detector, and, due to its nature, it is provided in a single transverse mode:

$$P_{LO} = h\nu \, \bar{n}_{LO} \, \Delta\nu_{LO} \qquad (6)$$

with $\bar{n}_{LO} \gg 1$, and its narrow linewidth $\Delta\nu_{LO} \ll \Delta\nu_S$ depending on the coherence length $L_{coh} = c/\Delta\nu_{LO}$.

## 2.2. Radiation noise

According to ref. [17], eqns. (9) to (12), the rms power fluctuations an antenna receives are in a single polarization

$$\overline{(\delta P)^2} = \overline{(\delta E_{total})^2}/(\Delta t)^2 = \Delta M_A \Delta M_L (h\nu)^2 \overline{\delta n^2}/(\Delta t)^2$$
$$= (h\nu)^2 \, 2\Delta M_A \, \Delta\nu \Delta f \, \overline{\delta n^2} \qquad (7)$$

wherein the mode occupation number fluctuations $\overline{\delta n^2}$ are determined in the following.

### 2.2.1. Quantum noise propagation

After photons have been generated, they eventually reach the detector and are converted into photoelectrons. On this way they undergo several processes through which they can be deleted. Quantum-mechanically, this introduces additional noise. Calculation of quantum noise propagation after the Burgess variance theorem [42] gives for the noise propagation through a stochastic particle deletion process, i.e. attenuation through a medium, transfer to another mode, or conversion to photoelectrons during detection, with survival efficiency $\eta$:

$$\boldsymbol{\delta n'} = \sqrt{\eta(1-\eta)\bar{n}} \cdot \hat{\boldsymbol{n}}_\eta + \eta \cdot \boldsymbol{\delta n} \quad (8)$$

with $\boldsymbol{\delta n} = \delta n \cdot \hat{\boldsymbol{n}}$, and $\boldsymbol{\delta n'} = \delta n' \cdot \hat{\boldsymbol{n}}'$, where the $\hat{\boldsymbol{n}}$ noise phasors are fast-changing complex Gaussian (2D-bell shaped) random variables, see discussion in [17]. $\bar{n}$ is the photon number per mode before deletion and $\bar{n}' = \eta\bar{n}$ is the photon number after deletion. In HD we may interpret here $\bar{n}'$ also as the IF photon number in a mode of the IF circuit (parametric down-conversion process, detection not yet accomplished). $\hat{\boldsymbol{n}}_\eta$ is the stochastically independent complex deletion noise phasor, so that $\overline{\hat{\boldsymbol{n}}_\eta \cdot \hat{\boldsymbol{n}}} = 0$, resulting for the propagated photon number $n'$ in

$$\overline{\delta n'^2} = \eta(1-\eta)\bar{n} + \eta^2 \cdot \overline{\delta n^2} \quad (9)$$

The first term on the right sides of eqns. (8) and (9) represents independent noise newly generated from stochastic deletions (which describes, for example, independent new shot noise generated inside a detector). The second term represents the transferred (i.e. detected) original radiation noise. With $\overline{\delta n^2} = \bar{n}$ eq. (9) shows also that the shot noise level of a laser is preserved through photo-electric detection with efficiency $\eta$, i.e. $\overline{\delta n'^2} = \bar{n}'$ where $\bar{n}' = \eta\bar{n}$, or through any attenuation. Eqns. (8) and (9) have worked well for explaining the observed active reduction of laser excess noise in a microwave photonic circuit over a bandwidth of several GHz to near the shot-noise limit [43].

### 2.2.2. Thermal radiation noise

The variance of the semi-classical thermal mode occupation number in eqn. (4) (Planck formula) is given by [1]

$$\overline{(\delta n)^2} = \bar{n}(\bar{n} + 1) \quad (10)$$

Detecting thermal radiation noise with efficiency $\eta$ results according to eqn. (9) in $\overline{\delta n'^2} = \eta\bar{n}(\eta\bar{n} + 1)$, a relation found in literature, but without derivation [23] [44]. For the attenuation in the atmosphere or in the telescope optics the identification $\eta \equiv \mathcal{T}$ must hold.

$$\overline{\delta n_S^2} = \eta_S \mathcal{T}\bar{n}_S(\eta_S \mathcal{T}\bar{n}_S + 1) \quad (11)$$

But how noise enters due to the thermal foreground emission? Unfortunately, eqn. (9) was derived with accounting only for stochastic deletion of photons and not for their stochastic generation. But if the emission process is seen as a transmission probability from the inside of the emitter to its outside (into the radiation mode), then it must hold $\overline{(\delta n_B)^2} = \mathcal{E}\bar{n}_B(\mathcal{E}\bar{n}_B + 1)$ and because of $\mathcal{E} = \mathcal{A} = 1 - \mathcal{T}$ we get:

$$\overline{\delta n_B^2} = (1-\mathcal{T})\bar{n}_B\big((1-\mathcal{T})\bar{n}_B + 1\big) \quad (12)$$

The thermal noise due to the signal S in a single mode and polarization is:

$$\overline{\delta P_S^2} = (h\nu)^2 \Delta\nu\, 2\Delta f\, \mathcal{T}\eta_S\bar{n}_S(\mathcal{T}\eta_S\bar{n}_S + 1) \quad (13)$$
$$= 2h\nu\Delta f\,(\mathcal{T}\eta_S\bar{n}_S + 1)\cdot P_S$$

using eqn. (2b). The background B is given according to eqn. (7) by (for a single polarization and $\Delta M_A$ modes):

$$\overline{\delta P_B^2} = (h\nu)^2 \Delta\nu\, 2\Delta f\, \Delta M_A(1-\mathcal{T})\bar{n}_B\big((1-\mathcal{T})\bar{n}_B + 1\big) \quad (14)$$
$$= 2h\nu\Delta f\big((1-\mathcal{T})\bar{n}_B + 1\big)\cdot P_B$$

In DD the detector pixels are sensitive to many modes, and so the form of eqn. (14) would be more generally:

$$\overline{\delta P_{S,DD}^2} = (h\nu)^2 \Delta\nu\, 2\Delta f\, \mathcal{T} \sum_M \eta_{S,M}\bar{n}_S(\mathcal{T}\eta_{S,M}\bar{n}_S + 1) \quad (15)$$

The different conventions in noise characterization are explained in the appendix 8.1.1, i.e. the NEP versus the System Noise Temperature, from which results also an alternative derivation of the traditional SNR formula in radio astronomy.

### 2.2.3. Laser noise

Now we add laser noise for the case of HD. In the semi-classical picture, sufficient for the description of the photon detection process [45], the minimum possible variations of the coherent laser field at high photon occupation numbers (minimum phase fluctuations according to the uncertainty principle $\Delta\varphi = 1/2\Delta n$) are governed by the Poissonian probability distribution (shot noise limit) with variance $(\delta n_{LO})^2 = \bar{n}_{LO}$ (both $\gg 1$) in a single mode arriving at a fast (coherent) detector. It is assumed that we can approximate any white amplified spontaneous emission (ASE) excess noise by

$$\overline{(\delta n_{LO})^2} = F\,\bar{n}_{LO} \quad (16a)$$

with $F := \overline{(\delta n_{LO})^2}/\bar{n}_{LO} \geq 1$ the Fano-factor [43]. $F = 1$ means that the laser operates at the shot noise limit. With the time scale $\Delta t = 1/2\Delta f$, an equivalent expression on the radiation side is

$$\overline{(\delta P_{LO})^2} = 2F\,h\nu\bar{P}_{LO}\Delta f \quad (16b)$$

We note here that in HD SNR calculations the quantum limit of these laser fluctuations ($F = 1$) is equivalent (has the same effect) as the zero-point fluctuations (ZPF). In this way, the classical derivation of the post-detection SNR of heterodyne (see appendix 8.1.4), the shot-noise of the detector is assumed as the only noise and is backwards reinterpreted as the ZPF determining the HD NEP. Theoretically, a Fano-factor of $F = 1$ can also be achieved in HD by using a balanced mixer which eliminates the LO excess noise, but in reality, only $F = 2 \cdots 4$ can be reached.

## 3. Signal-to-Noise Ratios (SNRs)

It makes sense to define the pre-detection (radiation-noise limited) signal-to-noise ratio (SNR) in a quadratic manner as

$$SNR_{pre} := \bar{P}_{rad}^2 / \overline{\delta P_{rad}^2} \quad (17)$$

In the further discussion we consider for $\overline{\delta P_{rad}^2}$ the variations resulting from comparing many subsequent single temporal modes ($M_L = 1$), which represent the shortest possible, independent measurement integrations, although technically hardly achievable for channel widths larger a few megahertz. The integration time corresponding to a single temporal mode is then the bandwidth coherence time, see eqn. (1d).

### 3.1. Post-detection SNRs

After detection and integration (low-pass filtering), an arbitrary integration time is then realized by averaging over $M_L = \Delta t/t_{coh} = \Delta\nu\Delta t = \Delta\nu/2\Delta f$ temporal modes which will increase the SNR by the factor $\sqrt{M_L} = \sqrt{\Delta\nu\Delta t}$ (see e.g. Wilson et al. [39]) for both detection methods equally.

Also, several deteriorating factors appear, like the quantum efficiency $\eta_Q < 1$ of the detector, and any post-detection electronic noise $N_{el}$, as that from the post-detection amplifier projected to the detector, and from dark current through the detector.

For photoconductive and photon-counting detectors it holds $I_{el,rad} = \Re P_{rad}$, where $\Re = \eta_Q e/h\nu$ is the responsivity. With $\overline{P_{el,rad}} = Z\overline{|I_{el,rad}|^2}$ (single receiver/detector) this leads to

$$\overline{P_{el,rad}} = Z\Re^2 \overline{P_{rad}^2} \tag{18}$$

where $Z$ is the impedance of the transmission lines (usually 50 Ω). Accordingly, electronic radiation noise power is given as

$$N_{el,rad} = Z\overline{|\delta I_{el,rad}|^2} = Z\Re^2 \overline{(\delta P_{rad})^2} \tag{19}$$

and purely electronic noise power adds to this, $N_{el,tot} = N_{el,rad} + N_{el}$.

This way, eqn. (17) is consistent with the post-detection (electronic) signal-to-noise ratio being linear in the post-detection electric power due to radiation absorption:

$$SNR_{post} = \overline{P_{el,rad}}/N_{el,tot} \tag{20}$$

Eq. (20) is equal to $\overline{I_{el,rad}^2}/\overline{\delta I_{el,tot}^2}$ in case of all post-detection impedances being the same. Even for slow direct detectors, i.e. bolometers and CCDs, this square law relation holds, since $\Delta I \propto \Delta P_{rad}$, but there is the choice of amplifying the current change directly, e.g. in a transimpedance amplifier, and so it would make sense to regard also $\widetilde{SNR}_{post} := \overline{I_{el,rad}}/\overline{|\delta I_{el,tot}|}$.

Whereas in DD the spectral separation of the light and superposition of the beams from the different telescopes of an interferometer ("fringes") is performed optically before detection (over an array of many detectors), in HD it is done after down-conversion into the IF (in basically a single-detector, or two balanced) by means of electronic filters and correlators. The modern approach is to use digital auto-correlators (AC) to determine the IF-spectrum from a single-receiver and cross-correlators (CC) for measuring the fringes between any two of the telescopes, where both types of correlations are performed alongside each other in one and the same digital platform (e.g. in a digital circuit programmed into an FPGA). In those calculations the expressions for the AC and CC IF-powers have the same structural form, i.e.

$$\overline{P_{IF,AC,i}(\omega)} = \frac{1}{Z}\overline{\widetilde{V}_i(\omega)\widetilde{V}_i^*(\omega)} = \frac{1}{Z}\overline{|\widetilde{V}_i(\omega)|^2} \tag{21a}$$

$$\overline{P_{IF,CC,i,j}(\omega)} = \frac{1}{Z}\overline{\widetilde{V}_i(\omega)\widetilde{V}_j^*(\omega)} \tag{21b}$$

where $\widetilde{V}_i(\omega) = ZI_{el,i,rad}$ are the post-detection voltage parts due to the radiation (which are proportional to the signal E-fields, which contain in turn the radiation noise) plus the noise voltage from the receiver noise. These two equations state that, if $SNR_{AC}$ has a form proportional to the optical input power, this must also hold for $SNR_{CC}$. Especially, forms like $SNR_{CC} \propto SNR_{AC}^2$ must be wrong, which is shown explicitly in the appendix 8.3 and was experimentally verified.

In the following, we first derive how the post-detection SNR results from the pre-detection SNR. Then we determine the expressions for pre-detection photon-noise limited HD and DD SNRs assuming the previously derived photon statistics in single temporal modes.

Considering averages and standard deviations over single temporal modes, $\overline{(...)^1}$, we get

$$SNR_{post}^1 = \frac{\overline{P_{el,rad}^1}}{N_{el,tot}^1} \approx \frac{Z\Re^2(\overline{P_{rad}})^2}{Z\Re^2\overline{(\delta P_{rad}^1)^2} + N_{el}^1} = \frac{SNR_{pre}^1}{1+NR^1} \tag{22}$$

with the noise ratio defined as

$$NR^1 := \frac{N_{el}^1}{Z\Re^2\overline{(\delta P_{rad}^1)^2}} \tag{23}$$

Averaged over the number of temporal modes $M_L = \Delta\nu\Delta t \geq 1$, each of them being a statistically independent measurement time, the SNR increases by $\sqrt{M_L}$, so that the final post-detection SNR is

$$SNR_{post} = \frac{SNR_{pre}^1}{1+NR^1}\sqrt{\Delta\nu\Delta t} \tag{24}$$

While $SNR_{pre}^1$ tends to be $< 1$ as we will see, only by multiplication with the large factor $\sqrt{\Delta\nu_S\Delta t}$ the $SNR_{post}$ becomes larger than unity.

For the two detection schemes, this integration takes place at different stages, as illustrated in Fig. 1. In HD the beat-photoelectrons are alternating-current (ac) which is amplified before it is rectified and integrated according to eqns. (21). We may view HD as a down-conversion of photons which then appear as quanta of the IF-mode from which they are then square-law-detected. In DD the photoelectrons are decent-current (dc), in principle directly squaring and integrating the source's E-field, before they are read out and amplified. Here, eq. (24) is valid for each pixel separately before readout, while the distribution of the signal power into m pixels leads to the fact that readout further reduces the SNR as is derived in 3.3.2. and 8.4.

As derived in the appendix 8.2., the *NR* for HD is

$$NR_{HD}^1 \approx \frac{1}{(1-\mathcal{T})\bar{n}_B + 2}\left(\frac{1}{\eta_Q} - 1\right) \tag{25a}$$

whereas for DD it is

$$NR_{DD}^1 \approx \frac{1}{(1-\mathcal{T})\bar{n}_B + 1}\left(\frac{1}{\eta_Q}\left(1 + \frac{N_{dark}}{N_{rad}^1}\right) - 1\right) \tag{25b}$$

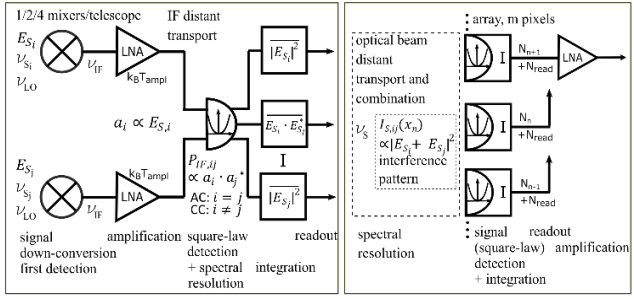

Fig. 1: Clarifying the ordering of square-law detection, low-noise amplification (LNA), integration (I) and readout for HD (left) and DD (right) interferometry.

where $N_{rad}^1 = (\mathcal{T}\eta_S \bar{n}_S + (1-\mathcal{T})\Delta M_A \bar{n}_B)$ according to eqn. (6), which is dominated usually (for $\eta_S \ll 1$) by the background.

In case the background is very low (in the near-IR/vis, and in case of spaceborne systems also for the mid-IR and far-IR), together with a high spectral resolution, $N_{rad}$ can become smaller than $N_{dark}$ (to which we count also the readout-noise electrons $N_{read}$), so that $NR_{DD}^1$ diverges. For ground-based observations, the ratio $N_{dark}/N_{rad}^1$ is more easily diminished than in space. The thermal sky background at 10 μm wavelength is as high as $B_{sky} = 100 \frac{Jy}{as^2} = 4.25 \cdot 10^{-14}$ WHz$^{-1}$m$^{-2}$Sr$^{-1}$ [47], so that the ratio $N_{dark}/N_{rad}^1$ is comparable to unity only for the smallest integration times, e.g. for $\Delta t = 1$ ms, which is used at Paranal also in the H and K bands to not degrade the visibility, see appendix eqns. (A44)-(A46), and smallest bandwidths (e.g. for R > 10000), since we obtain

$$N_{rad} \gtrsim \frac{B_{Sky}}{h} \cdot \lambda^2 \cdot \frac{\Delta t}{R} \qquad (26)$$

independently of the telescope because of $\Delta\Omega\Delta A_T \geq \lambda^2$. But for space-borne DD interferometry in the mid-infrared range (10 − 20 μm), the divergence of *NR* towards low intensities should become very relevant, since $B_{sky}$ is 4-5 orders of magnitude less in the direction of the zodiacal light [48], and else even less. Therefore, $N_{read}$ of future mid-IR direct detectors is desirable to be reduced by orders of magnitude, which is not yet in sight.

In the following two sections, we derive the pre-detection $SNR_{pre}^1$ expressions to be inserted into eqn. (24). "Pre" means here that we do not consider the non-ideality of detection, but we do take already into account the principle of the detection, as splitting to balanced mixers in HD or distribution over many pixels in DD.

### 3.2. HD pre-detection SNR

In the following two sub-sections, we derive $SNR'_{pre}$ for a single-telescope receiver and then for single-baseline interferometry, i.e. cross-correlation (CC), and we will see that the latter is in complete analogy to auto-correlation, in particular it does not result in another power law. The total resulting instantaneous radiation power before detection, neglecting products without the LO-field in the limit $P_{LO} \gg P_S, P_B$, is at each telescope

$$P_{total}(t) = Ac\varepsilon_0 \overline{|E_S(t) + E_B(t) + E_{LO}(t)|^2} \qquad (27a)$$

$$\approx P_S + P_B + P_{LO} + P_{het}(t)$$

where the average is over several periods of *S* or *LO*, with

$$P_{het}(t) = 2\sqrt{P_S P_{LO}} \cdot \sin(\omega_{IF}t + \varphi_S(t)) + 2\sqrt{P_B P_{LO}} \\ \cdot \sin(\omega_{IF}t + \varphi_B(t)) \qquad (27b)$$

because the laser is operating single-mode, which is matched to the receiver optics, is polarized, and picks out only one spatial mode and polarization of the thermal signal.

After detection we have $P_{IF} = P_{el} = ZI_{el}^2 = Z\Re^2(P_{het})^2 = (2Z\Re^2 P_{LO})P_S$, and $I_{el} \propto E_S$. The heterodyne signal-phasor can be written with rms-averaged amplitudes

$$\widehat{P}_{het} = \sqrt{2P_S P_{LO}} e^{i\varphi_S(t)} + \sqrt{2P_B P_{LO}} e^{i\varphi_B(t)} \qquad (27c)$$

The unwanted thermal background of the atmosphere and the telescope optics is indistinguishable from the thermal signal *S* if looking at only one angular sky position. We assume that the ZPFs can be regarded as included in the mode-filling thermal background, which is at $T_B > 0$. As the phases of both, the signal and the background heterodyne beatings, are uncorrelated, the rms-square of the heterodyne and background power signals are simply adding up:

$$\overline{(P_{het})^2} = 2P_S P_{LO} + 2P_B P_{LO} \qquad (27d)$$

$2P_B P_{LO}$ is undergoing only slow atmospheric drifts and should not vary over small enough portions of the sky, so that it can be subtracted from the total signal at the position of the point source by position-switching quickly enough between the source (ON) and a position on the sky with negligible astronomical intensity or a load (OFF) with duty cycle 1:1 (ON/OFF-integration scheme). This is also true for DD, but in heterodyne, in case of emission or absorption lines, frequency-switching the LO can be additionally applied. To account for this in the SNRs, the instantaneous heterodyne signal power phasor $\widehat{P}_{het,S}$ is divided by $\sqrt{2}$ in the following, rather than the noise terms multiplied by $\sqrt{2}$, see Dicke-switching principle as described in ref [39].

The upper and lower sideband of the signal can be represented by the conjugated complex parts:

$$\widehat{P}_{het,S}(\omega,t) = P_{het,USB}(\omega) e^{i(\omega t + \varphi_{S,USB}(t))} \\ + P_{het,LSB}(\omega) e^{-i(\omega t + \varphi_{S,LSB}(t))} \qquad (28)$$

In CC between two telescopes, both sidebands can be separated with phase-switching the LO by $\pi/2$ at one of the telescopes [11]. However, if possible it would be preferred to directly separate both sidebands in a balanced receiver using four mixers.

For one sideband and in a frame of the complex plane rotating with $\omega = \omega_{IF}$, we can write for the time-evolution of the IF Fourier-component at $\omega$:

$$\widehat{P}_{het}(\omega,t) = \sqrt{P_S(\omega)P_{LO}}\,\widehat{s}(\omega,t) + \delta P_S(\omega)\widehat{n}_S(\omega,t) \\ + \delta P_B(\omega)\widehat{n}_B(\omega,t) \\ + \delta P_{LO}(\omega)\widehat{n}_{LO}(\omega,t) \qquad (29)$$

$\widehat{s}(\omega,t) = \widehat{s}_0 e^{i\varphi_S(t)}$ is the normalized IF signal phasor. $\varphi(t)$ contains, besides the atmospheric phase perturbations and LO-phase drifts, also a phase-leftover from subtracting out the slowly varying thermal background with the Dicke-switching-principle.

The power fluctuations from eqns. (13), (14) and (16b) are phasors:

$$\delta \widehat{P}_i(t) = \sqrt{\overline{\delta P_i^2}} \cdot \widehat{n}_i(t) =: \delta P_i \cdot \widehat{n}_i(t) \quad (30)$$

where $\widehat{n}_i(t)$ ($i = S, B, LO$) are fast-changing complex Gaussian (2D-bell shaped) random variables, as introduced in eqn. (8). They describe equal probabilities for all phase angles and Gaussian probability distribution for the amplitude, and are normalized and uncorrelated, i.e. $\int \widehat{n}_i(t) \cdot \widehat{n}_j(t) dt = \delta_{ij}$ (orthonormal). We are doing this to treat AC and CC in the same way. Therein, the $\delta P_i$ are expressed as:

$$\overline{|\delta P_S|^2} = 2h\nu\Delta f \, (\mathcal{T}\eta_S \bar{n}_S + 1) \cdot P_S \quad (31a)$$

$$\overline{|\delta P_B|^2} = 2h\nu\Delta f \, ((1-\mathcal{T})\bar{n}_B + 1) \cdot P_B \quad (31b)$$

$$\overline{|\delta P_{LO}|^2} = 2h\nu\Delta f \, F \cdot P_{LO} \quad (31c)$$

The absolute measure of radiation noise is much higher for the laser field than it is for the source and background fields, simply because of the much higher power. This might be the reason of a widespread belief that HD would be swamped by laser noise and be deteriorated compared to DD. But we show in the following sections by a semi-classical derivation, applied to one balanced mixer in AC and two balanced mixers in CC (a balanced correlation receiver, BCR, see Fig. 2), that the signal noise is amplified through the down-conversion by the huge factor of the ratio of laser power through the signal power, while the laser noise is reduced through this process by the inverse factor. This leads to the same result as Yuen and Chan [6] found fully quantum-mechanically, i.e. that the heterodyne noise is only determined by the signal noise and its quantum limit by the vacuum fluctuations in the signal mode.

### 3.2.1. Auto-correlation (AC) pre-detection SNR

To approach the quantum limit of heterodyne detection, balanced mixers were shown to be very efficient, since these mixers subtract out any directly detected fluctuations (dc- and slow ac-signals), especially those of the LO, because these come along in common-mode. Instead, they can only detect beat-notes between signal and LO (the down-converted heterodyne signal) because these result with a phase difference of 180° on both mixers.

Therefore, from now on we only consider the heterodyne down-conversion of all noise sources, i.e. their coherent detection through the mixing process. The converted noise power phasors should have a fixed phase-relation $\Delta\varphi_m$ with the original (directly detected) noise power phasors. While for a single-ended mixer it is

$$\delta P_{het} = \delta P_S \left(1 + e^{i\Delta\varphi_m} 2 \frac{\partial}{\partial P_S} \sqrt{P_{LO} P_S} \, e^{i\varphi_S}\right) + \quad (32)$$
$$\delta P_B \left(1 + e^{i\Delta\varphi_m} 2 \frac{\partial}{\partial P_B} \sqrt{P_{LO} P_B} \, e^{i\varphi_B}\right) +$$
$$\delta P_{LO} \left(1 + e^{i\Delta\varphi_m} 2 \frac{\partial}{\partial P_{LO}} \left(\sqrt{P_{LO} P_S} \, e^{i\varphi_S} + \sqrt{P_{LO} P_B} \, e^{i\varphi_B}\right)\right)$$

for balanced mixers, considering the subtraction of the directly detected noise of both mixers in the IF-band, and the 180° phase difference of the heterodyne down-converted noise at both mixers, it is:

$$\delta P_{het,BPD} := \sqrt{\frac{P_{LO}}{P_S}} \delta P_S \, e^{i\varphi_S} + \sqrt{\frac{P_{LO}}{P_B}} \delta P_B e^{i\varphi_B} \quad (33)$$
$$+ \frac{\sqrt{P_S} e^{i\varphi_S} + \sqrt{P_B} e^{i\varphi_B}}{\sqrt{P_{LO}}} \delta P_{LO}$$

$$\overline{|\delta P_{het,BPD}|^2} := \left(\frac{P_{LO}}{P_S} \overline{|\delta P_S|^2} + \frac{P_{LO}}{P_B} \overline{|\delta P_B|^2}\right) \quad (34)$$
$$+ \frac{P_S + P_B}{P_{LO}} \overline{|\delta P_{LO}|^2}$$

which uses that the phases $\varphi_S$ and $\varphi_B$ are uncorrelated. With eqns. (31), we get

$$\overline{|\delta P_{het,BPD}|^2} = 2h\nu\Delta f [(\mathcal{T}\eta_S \bar{n}_S + 1 + (1-\mathcal{T})\bar{n}_B \quad (35)$$
$$+ 1)P_{LO} + F(P_S + P_B)]$$

It follows the auto-correlation SNR

$$SNR_{AC} := \frac{(P_{het})^2}{\overline{|\delta P_{het,BPD}|^2}} \quad (36)$$

$$= \frac{\mathcal{T}\eta_S \, h\nu \, \bar{n}_S \, \Delta\nu_S \, P_{LO}}{2h\nu\Delta f [(\mathcal{T}\eta_S \bar{n}_S + 1 + (1-\mathcal{T})\bar{n}_B + 1)P_{LO} + F(P_S + P_B)]}$$

and in the limit of $P_{LO} \gg P_S, P_B$ and for one temporal mode ($2\Delta f = \Delta\nu_S$):

$$SNR_{AC}^1 \approx \frac{\mathcal{T}\eta_S \, \bar{n}_S}{(\mathcal{T}\eta_S \bar{n}_S + (1-\mathcal{T})\bar{n}_B + 2)} \approx \frac{P_{S,\nu}}{2h\nu} \quad (37)$$

This result is obviously determined by the thermal radiation noise of the signal rather than by the excess noise of the laser ($F > 1$), in agreement with the fully quantum-mechanical derivation of Yuen and Chan [6], which also was formulated for a balanced mixer which would achieve theoretically $F = 1$. In parallel to the above derivation, in the appendix 8.2.1 we reproduce the traditional result for a single mixer as appearing so far in the literature, which is determined by the laser noise featuring the Fano-factor $F$.

Because this factor is 1 for an ideal balanced mixer also with excess noise of the LO, but it is observed that even balanced mixer perform somewhat above the standard quantum limit (SQL), we introduce a heuristic/experimental factor $\mathcal{F} \geq 1$, describing the losses in a real balanced mixer not modelled here, pushing it away from the quantum limit, by replacing $SNR'_{AC} \rightarrow SNR'_{AC}/\mathcal{F}$ and $SNR'_{CC} \rightarrow SNR'_{CC}/\mathcal{F}$. In the plots we assume $\mathcal{F} = 3$, a value we reached at 1550nm [17] and others in the submm-range [17].

### 3.2.2. Cross-correlation (CC) pre-detection SNR

According to the Wiener-Khinchin theorem (see [39], p. 56-58, and its formalization for CC, e.g. in [17]) we can calculate the CC power spectrum between two telescopes after [11] p. 77, by calculating the fast-Fourier-transforms (FFT) of both single-receiver heterodyne voltage signals over a moving short time-window of length $T$

$$\tilde{V}_{het,j,T}(\omega,t) = \int_{t-T}^{t+T} dt' \, V_{S,j}(t') e^{i\omega t'} \quad (38)$$

multiplying them subsequently and integrating the product over a long time (FX-correlator). These post-detection expressions enter in eqn. (21b), and so we can write for pre-detection correspondingly:

$$\tilde{P}_{CC,T}(\omega,t) \propto \tilde{E}_{S1,T}(\omega,t) \cdot \tilde{E}_{S2,T}^{*}(\omega,t) \quad (39a)$$

and

$$\tilde{P}_{het,i,T}(\omega,t) \propto \tilde{E}_{S,i,T}(\omega,t) \cdot \tilde{E}_{LO,T}^{*}(t) \quad (39b)$$

With this we can write:

$$\overline{\tilde{P}_{het,1,T}(\omega,t) \cdot \tilde{P}_{het,2,T}(\omega,t)} = \overline{\tilde{P}_{CC,T}(\omega,t) \cdot P_{LO}} \quad (40)$$

which is analog to the auto-correlation expression of eq. (27c). Assuming equal strengths of signal and noise terms, respectively, at both telescopes, we have:

$$\overline{\tilde{P}_{het,1,T}(\omega,t) \cdot \tilde{P}_{het,2,T}^{*}(\omega,t)} = (S_{CC})^2 + \overline{(N_{CC}(t))^2} \quad (41a)$$

and according to eqns. (29) – (31), with the signal part

$$(S_{CC})^2 := P_S(\omega) P_{LO} \overline{\hat{\mathbf{s}}_1 \hat{\mathbf{s}}_2^{*}(\omega,t)} \quad (41b)$$

and the noise part

$$\overline{(N_{CC}(t))^2} := |\delta P_S|^2 \overline{\hat{\mathbf{n}}_{S1} \hat{\mathbf{n}}_{S2}^{*}(\omega,t)} \quad (41c)$$
$$+ |\delta P_B|^2 \overline{\hat{\mathbf{n}}_{B1} \hat{\mathbf{n}}_{B2}^{*}(\omega,t)} + |\delta P_{LO}|^2 \overline{\hat{\mathbf{n}}_{LO1} \hat{\mathbf{n}}_{LO2}^{*}(\omega,t)}$$

where the correlation terms between the random phasors (averaged over $T \ll \infty$ they are still time-dependent or not exactly zero) have the following meanings (see also [17]):

- $\overline{\hat{\mathbf{s}}_1 \hat{\mathbf{s}}_2^{*}} =: \gamma$    Normalized correlation of the signal waves (visibility), the quantity to be measured by an interferometer (the spatial coherence function of the signal waves over the projected baseline of the telescopes). (42a)

- $\overline{\hat{\mathbf{n}}_{S1} \hat{\mathbf{n}}_{S2}^{*}} =: c_S$    CC of the signal noise. According to eq. (8) it should hold $c_S = \eta_S$ for a spatial interferometer, where $\eta_S$ is the single-telescope beam-filling factor for the source as resolved by the interferometer. In case of a single-telescope balanced correlation receiver (BCR, lower part of Fig. 2) with a 50:50-splitter between the two bal. receivers it should be $c_S = c_{LO}$. (42b)

- $\overline{\hat{\mathbf{n}}_{B1} \hat{\mathbf{n}}_{B2}^{*}} =: c_B$    The thermal background at the two telescopes should be completely uncorrelated, i.e. $c_B = 0$, since in the near-field both telescopes look at different thermal emitter volumes (atmosphere) and surfaces (telescopes). However, if the background source is interstellar, then it should be counted to the source, i.e. $c_B = c_S$. (42c)

- $\overline{\hat{\mathbf{n}}_{LO1} \hat{\mathbf{n}}_{LO2}^{*}} =: c_{LO}$    correlation of the laser noise at both receivers. (42d)

This approach was also used in a report on the evaluation of visible light heterodyne interferometry [49], but its derivation therein contains a simple error as becomes clear in comparison to our derivation of the CC SNR in their eqns. (41)-(44): Where the $SNR_{zz*}$ is calculated, all the corresponding terms of eqns. (41b) and (41c) were entered quadratically, as if to calculate the term $(S_{CC}/N_{CC})^4$. Furthermore, the correlation coefficients we consider in eqn.

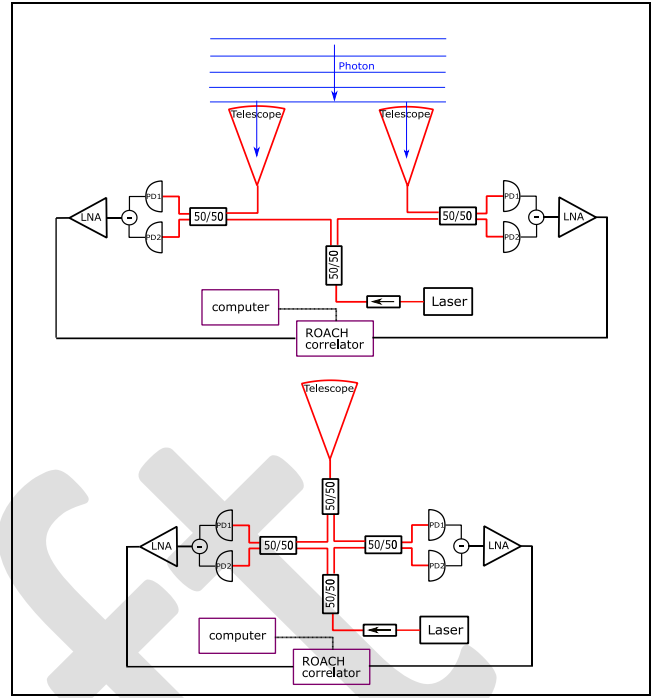

*Fig. 2: Top: Balanced Correlation Receiver (BCR) for heterodyne interferometry. Both telescopes look at the same transversal mode (the angular far-field pattern). A photon within the coherence area of the interferometer (see e.g. [50]) enters any of the telescopes with the probability $\eta_{S,,}$. Therefore, both probabilities are not correlated. Bottom: BCR for single telescopes. Here a photon from the telescope or from the LO undergoes two independent "decisions" in series before it hits one of the detectors, and therefore also here the four probabilities, or the two differences of them, are not correlated.*

(41c) were all assumed to be unity, except the signal correlation, which was equally assumed as the visibility. Avoiding this mistake, we arrive at an equally linear expression in $P(\nu)$ for the CC-*SNR*, which is consistent with the linear plots of our laboratory noise temperature measurements with two balanced receiver systems at $\lambda = 1550$ nm [17], and anyway clear from eqns. (21a) and (21b).

Because $\overline{|\delta \mathbf{P}_{het,A}|^2} = \overline{|\delta \mathbf{P}_{het} \cdot \delta \mathbf{P}_{het}^{*}|}$, we have for CC, analog to the AC-variance-term of eqn. (34):

$$\overline{|\delta \mathbf{P}_{het,BPD,A} \cdot \delta \mathbf{P}_{het,BPD,B}^{*}|} =$$

$$\left( \gamma c_S \frac{P_{LO}}{P_S} |\delta P_S|^2 + c_B \frac{P_{LO}}{P_B} |\delta P_B|^2 \right) + \frac{P_S + P_B}{P_{LO}} c_{LO} |\delta P_{LO}|^2$$

$$= 2h\nu \Delta f \big[ [c_S(\mathcal{T} \eta_S \bar{n}_S + 1) \quad (43)$$
$$+ c_B((1 - \mathcal{T}) \bar{n}_B + 1)] P_{LO}$$
$$+ F c_{LO}(P_S + P_B) \big]$$

$$SNR_{CC} = \frac{(P_{het,CC})^2}{\overline{|\delta \mathbf{P}_{het,BPD,A} \cdot \delta \mathbf{P}_{het,BPD,B}|}} \quad (44)$$

$$= \frac{\gamma P_S P_{LO}}{2h\nu \Delta f} \Big/ \left\{ \begin{matrix} [c_S(\mathcal{T} \eta_S \bar{n}_S + 1) + c_B((1 - \mathcal{T}) \bar{n}_B + 1)] P_{LO} \\ + F c_{LO}(P_S + P_B) \end{matrix} \right\}$$

For $P_{LO} \gg P_B, P_S$ and a single temporal mode it results:

$$SNR_{CC}^1 \approx \frac{\gamma \mathcal{T} \eta_S \bar{n}_S}{\eta_S(\mathcal{T}\eta_S\bar{n}_S + 1)} \quad (45)$$

$$\approx \left((1-\mathcal{T})\bar{n}_B + 2\right)\frac{\gamma}{\eta_S} \cdot SNR_{AC}^1$$

Therefore, the CC-SNR of two balanced mixers should be much better (by the large factor $1/\eta_S$, which is, however, expected to be limited by technical imperfections) than known from cross-correlating single mixers in two-telescope interferometry and in single-telescope correlation receivers [11]. There the LO-noise (and signal-noise in the latter) is not made uncorrelated at both receivers, and thus the CC-SNR is very similar to the AC-SNR of each receiver, because only the noise contributions of the post-detection IF-amplifiers cancel out from the output-noise but the LO-noise is preserved. In contrast to that, in the novel BCR receiver-architecture, the LO- and input-signal-noise both appear de-correlated in the IFs of the two balanced receivers, and so are also eliminated from the integrated CC quasi-DC output-signal. In our laboratory experiments at $\lambda = 1550$ nm we have obtained factors in the range of $c_{LO} = 0.1 \ldots 0.05$ using a SLED as an incoherent test-source [17], a result which was recently reproduced with a thermal light-source based on a high-pressure Hg-discharge lamp [18]. Furthermore, we obtained similar values for $c_{LO}$ using two balanced cryogenic superconducting junction (SIS) receivers in the submm-range at 460 GHz.

The dominant noise terms in eqn. (43) are the two stemming from the source- and background-noise, and not the one stemming from the LO-noise as it would result with ignoring the heterodyne down-conversion of all noise contributions. The latter procedure (noise seen only through direct detection, and in single-ended mixers) was so far used exclusively in the literature, see eqns. (A7) – (A11) of the appendix.

The use of single mixers instead of balanced ones can be represented in Fig. 2 with replacing the horizontal (receiver) 50/50-power-splitters by ones having e.g. 95% transmission for the signal and 5% for the LO, assuming that the LO has so much power that we can afford to discard most of it, but on the other hand we almost don't have losses for the signal path of one mixer (scheme used in most submm-receivers nowadays). This way, one of the two balanced mixers in each receiver becomes superfluous and can be taken out. Now a source photon is not anymore undergoing two "decisions" in which mixer to be detected, but effectively just one. Therefore, the photon-noise of the source, having been shown above as dominating the heterodyne signal noise, is now not anymore uncorrelated in the IFs of both receivers, but correlated with a minus-sign.

The question arising then within this theory would be what happens if we take a high-number-output splitter for the LO-distribution and use two of its outputs to feed the two mixers considered above. This behavior has to be measured still. The alternative would be two phase-locked LOs with uncorrelated proper noise. However, then both LO-noises and both mixers should be as nearest as possible at the quantum limit, since no balancing would be active. In the submm-range, cryogenic single mixers with a noise temperature of two times the quantum limit were demonstrated [51].

### 3.3. DD pre-detection SNRs
#### 3.3.1. Single-telescope DD SNR

The pre-detection signal-to-noise ratio in DD is for a polarized source, detected on one pixel, using eqns. (5), (8) and (9):

$$SNR_{DD,1px} := \frac{P_S^2}{|\delta P_S|^2 + |\delta P_B|^2} \quad (46)$$

$$= \frac{(\mathcal{T}\eta_S \bar{n}_S \Delta\nu_S)^2/(\Delta\nu_S 2\Delta f)}{[\mathcal{T}\eta_S\bar{n}_S(\mathcal{T}\eta_S\bar{n}_S + 1) + \Delta M_A (1-\mathcal{T})\bar{n}_B((1-\mathcal{T})\bar{n}_B + 1)]}$$

In case of very high transmission, $\mathcal{T} \to 1$, and low background, this reduces for a single temporal mode to

$$SNR_{DD,1px}^1 \approx \frac{\mathcal{T}\eta_S\bar{n}_S}{(\mathcal{T}\eta_S\bar{n}_S + 1)} \approx \frac{P_\nu}{h\nu} \quad (47a)$$

Note that this has the same form regarding the quantum noise as for HD in eq. (37).

…..

In case of the signal power very weak against the background power, $\mathcal{T}\eta_S\bar{n}_S \ll \Delta M_A(1-\mathcal{T})\bar{n}_B$, the functional form changes over to quadratic:

$$SNR_{DD,1px}^1 \approx \frac{(\mathcal{T}\eta_S\bar{n}_S)^2}{\Delta M_A(1-\mathcal{T})\bar{n}_B((1-\mathcal{T})\bar{n}_B + 1)} \quad (47b)$$

leading to a larger detection limit due to the small filling factors typical for interferometry, see Fig. 7. This can even be the case in space, with zodiacal light as a background, if looking at signals at the detection limit.

#### 3.3.2. Interferometric DD SNR

In a DD Michelson-type stellar interferometer, the light beams from two or more telescopes are superimposed on the same imaging detector (e.g. CCD, EMCCD, or APD-array) to form fringes across the star image. Assuming no atmospheric perturbations, the latter is the single-telescope PSF, the circular symmetric Airy pattern $I_A(\varphi, \vartheta)$ of angular width $\theta_B = 1.22\lambda/D$, with $D$ the telescope diameter. Therefore, the on-sky beam-pattern of the two-telescope interferometer has the form:

$$I_{int}(\varphi) = 2I_A(\varphi, \vartheta)\left[1 + V_{pre} \cos\left(2\pi \frac{b}{\lambda}\varphi\right)\right] \quad (48)$$

The many sensitivity maxima (sub-lobes), evenly spaced in $\varphi$-direction over the single-telescope lobe, are separated by zero-sensitivity points (destructive interference between both telescopes), having the angular separation $\Delta\varphi = \lambda/b$, in which $b \gg D$ is the spatial separation of the telescopes (baseline).

As the thermal background intensity from the atmosphere and the telescope optics is uncorrelated between the telescopes, it adds up in the interference pattern to a high background and therefore contrast is lost on the interference fringes. The background can be subtracted out by position switching and phase modulation, but its radiation noise adds up anyway.

Different from HD, in DD interferometry there is no way to measure the interference pattern with concentrating the source power to a single detector. Rather, the same total power must be diluted to $m$ parallel detectors (pixels) to determine the visibility and its phase. (An alternative would be $m$ measurement time intervals for using one detector at different OPDs between the telescopes, but this has the disadvantage of losing integration time and of being more affected by drifts.)

| | for both detection schemes: |  |
|---|---|---|
| $\eta_S \ll 1$<br>$\Delta M_A \gtrsim 1$<br>$\mathcal{T} < 1$ | $SNR_{post} = \dfrac{SNR^1_{pre}}{1 + NR^1}\sqrt{\Delta\nu \Delta t}$ | (24) |
| | to be inserted into eqn (24): | |
| noise reduction factor | for heterodyne detection (HD):<br>$NR^1_{HD} \approx \dfrac{1}{(1-\mathcal{T})\bar{n}_B + 2}\left(\dfrac{1}{\eta_Q} - 1\right)$    (25a)/(A16a) | for direct detection (DD):<br>$NR^1_{DD} \approx \dfrac{1}{(1-\mathcal{T})\bar{n}_B + 1}\left(\dfrac{1}{\eta_Q}\left(1 + \dfrac{N_{dark}}{N^1_{rad}}\right) - 1\right)$   (25b)/(A16b)<br>$N^1_{rad} = \mathcal{T}\eta_S\bar{n}_S + (1-\mathcal{T})\Delta M_A \bar{n}_B$    (6) |
| only in CC | for HD and DD, only for CC: insert $SNR_{post}$ of eqn. (24) into $SNR_{CC}$<br>$SNR_{CC,\varphi} = SNR_{CC}\, e^{-1/SNR_{CC}}$    (A46) | |
| enter for $SNR^1_{pre}$ | for a single telescope (AC): | for two-telescope interferometer ($\gamma = V_{pre}$) (CC): |
| for HD: | $SNR^1_{AC,HD} = \dfrac{\mathcal{T}\eta_S \bar{n}_S}{(\mathcal{T}\eta_S\bar{n}_S + (1-\mathcal{T})\bar{n}_B + 2)}$   (37) | $SNR^1_{CC,HD} \approx \dfrac{\gamma \mathcal{T}\eta_S \bar{n}_S}{c_S(\mathcal{T}\eta_S\bar{n}_S + 1)} \approx 2\dfrac{\gamma}{\eta_S}SNR^1_{AC,HD}$   (45) |
| for DD: | $SNR^1_{DD,1px} = $    (46)<br>$\dfrac{(\mathcal{T}\eta_S\bar{n}_S)^2}{[\mathcal{T}\eta_S\bar{n}_S(\mathcal{T}\eta_S\bar{n}_S + 1) + \Delta M_A(1-\mathcal{T})\bar{n}_B((1-\mathcal{T})\bar{n}_B + 1)]}$ | $SNR_{post,V,mpx}$<br>$= V_{pre}\sqrt{\dfrac{N_{rad} + N_{read}}{N_{rad} + mN_{read}}}\,SNR_{post,1px}$    (49)/(A36)<br>$N_{rad} = N^1_{rad}\Delta\nu_S\,\Delta t$ |

Table 1: Summary of the results of sections 3.2, 3.3 and the appendix to be entered into eq. (24).

For interferograms imaged separately for each baseline at least $m = 4$ must be used, which is the case in the GRAVITY-instrument [38] which employs integrated waveguide optics. To resolve the interference fringes of all baselines superimposed on one pixel-array, AMBER used 30 pixels for 3 baselines [52] and MATISSE uses 72 pixels for 6 baselines [16].

In the appendix 8.4 it is evaluated what this dilution over multiple pixels means for the SNR. It results the following dependence on the pixel number $m$

$$SNR_{V,mpx,post} = V_{pre}\sqrt{\frac{N_{rad} + N_{read}}{N_{rad} + mN_{read}}}\,SNR_{1px,post} \quad (49)$$

in which $N_{rad} = \eta_Q(\mathcal{T}\eta_S\bar{n}_S + (1-\mathcal{T})\Delta M_A\bar{n}_B)\,\Delta\nu_S\,\Delta t$ is the number of photoelectrons, according to eqn. (6), and $N_{read} = \overline{(\delta N_{read})^2}$ is the number of read-noise electrons. The effect of this is illustrated in Fig. 3.

In the visible and NIR-range, where the background is low, noiseless readout has been developed with electron avalanche-multiplication CCDs (EMCCDs, $N_{read} < 1$), and with e-APDs up to around $\lambda = 3\,\mu m$ [53]. Compared to this, the situation at $\lambda = 10\,\mu m$ appears rather dramatic: The cryogenic Aquarius detector of the new MATISSE instrument, operating in the N-band (8-13 μm), is limited by $\overline{(\delta N_{read})^2} = N_{read} \approx 9\cdot 10^4$) [46].

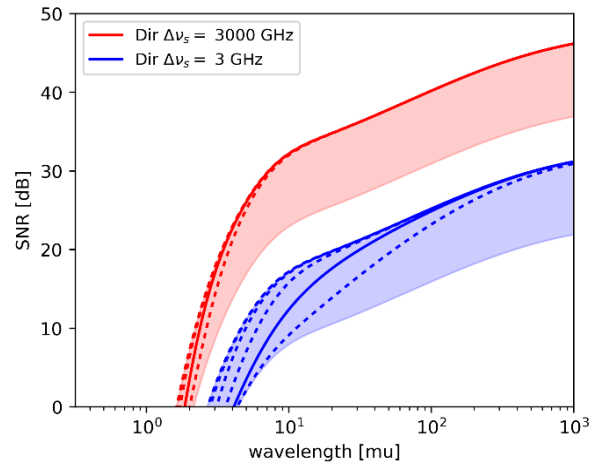

Fig. 3: The deteriorating influence of different (constant) $N_{read}$-values (dashed curves, from right to left: $N_{read} = 10^6, 10^5$ (full curve, used later), $10^4, 10^3, 10^2$) on the post-detection SNR for the example of $m = 72$, $\mathcal{T} = 0.9$, $\eta_S = 10^{-2}$, $T_S = 1000\,K$, $T_B = 300\,K$. The upper (red) curve-band is for $\Delta\nu_S = 3\,THz$, and the lower (blue) is for $\Delta\nu_S = 3\,GHz$, wherein the upper bounds are $SNR_{DD,1px,post}$-curves as in Fig. 4 (single telescope), and the lower bounds are the limits $SNR_{DD,1px,post}/\sqrt{m}$.

While in the K-band, HgCdTe-based avalanche photodiode (eAPD) arrays can reach an effective read noise of <1 e⁻ rms [38], the Aquarius detector of the MATISSE instrument, sensitive around 10 microns, is state-of-the-art with $N_{read} = 90000$ electrons per pixel [16]. This is incorporated into the simulations with a transition of $N_{read}$ from 0 to $10^5$ centered at 5 μm.

## 3.4. Comparison of the SNRs between HD and DD

The resulting equations for the pre-detection SNRs to be inserted into eq. (24) to obtain the post-detection SNRs are summarized in Table 1. Those are implemented into several Python routines to generate the example plots in the figures 4-7 without any approximations. In Fig. 4 we plot the post-detection SNRs for single telescopes in dependence of the wavelength, and in Fig. 5 for two-telescope interferometers, each for two different beam-filling factors. In Fig. 7 we plot the SNRs as a function of source power for single telescopes and two-telescope interferometers for the wavelengths $\lambda = 10\ \mu m$ and $\lambda = 20\ \mu m$. In Fig. 6 we plot the ratios of the HD-SNR to the DD-SNR over the wavelength.

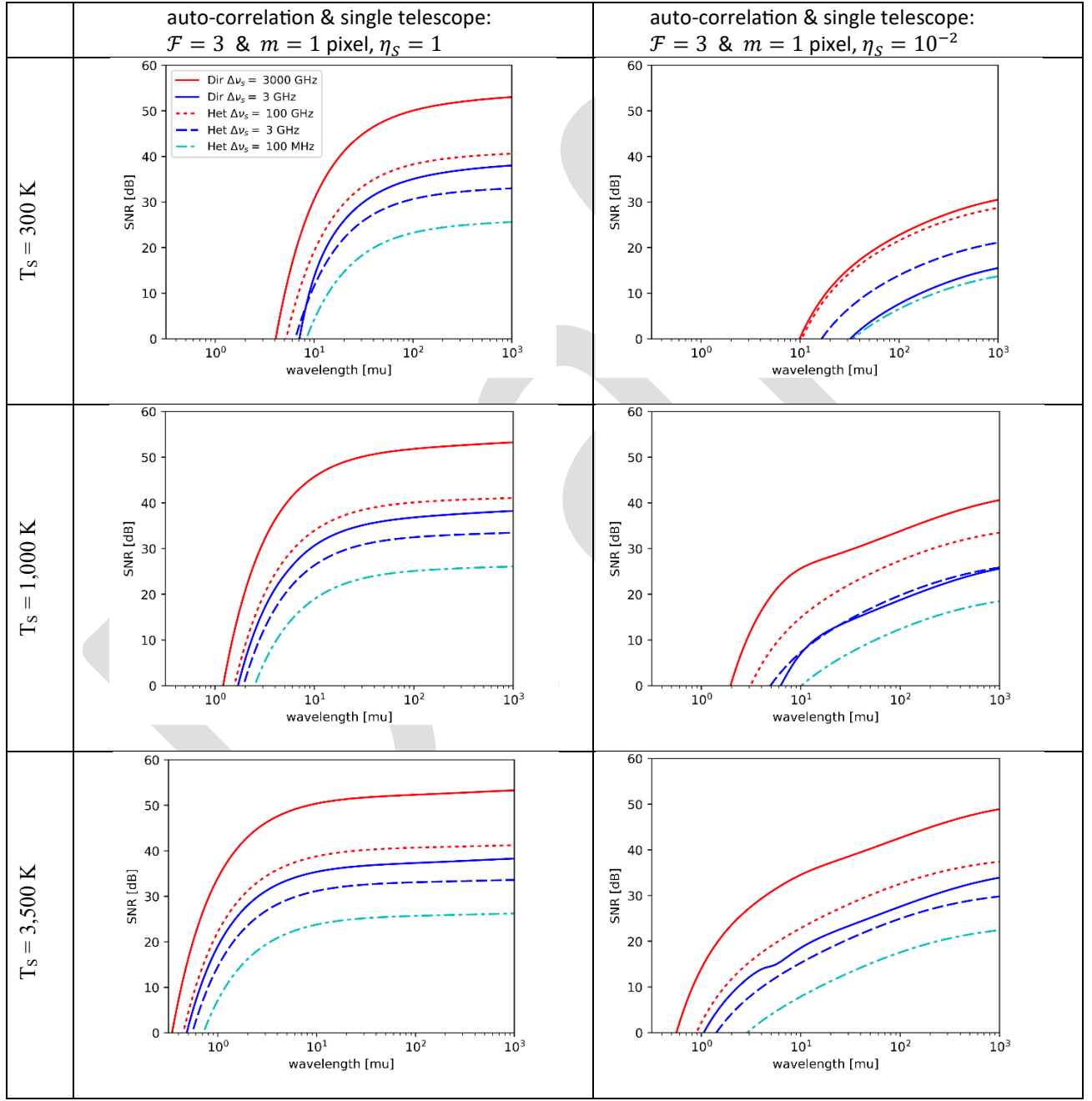

*Fig. 4: Single-telescope post-detection SNR comparison over wavelength for a single pixel/receiver, with the parameters $T_B = 300\ K$, $\Delta M_A = 1$, $\mathcal{T} = 0.9$, and $\Delta t = 10\ ms$. HD (eqn. (24c), channel bandwidths $\Delta \nu_S = 100\ MHz$, 3 GHz and 100 GHz) versus DD with 1 pixel (eqn. (27a), $\Delta \nu_S = 3\ GHz$ and 3 THz), for $\eta_S = 1$ (left panel) and $\eta_S = 0.01$ (right panel), and astronomical source temperature a) $T_S = 300\ K$ (dust in accretion disk), b) 1,000 K (young hot planet), and c) 3,500 K (red giant or dwarf). The blue curves can be compared directly.*

As common parameters we use $T_B = 300$ K, $\Delta M_A = 1$, $\mathcal{T} = 0.9$. The integration time $\Delta t = 10$ ms (in the plots over $\lambda$), is assumed as the best atmospheric coherence time at $\lambda = 2\ \mu m$ (K-band) at Paranal, where, however, the fringe integration times in interferometry often must be reduced to 1 ms, especially at the UT-telescopes. On the other side, the coherence time can be as long as 60 ms at $\lambda = 10\ \mu m$ and 140 ms at $\lambda = 20\ \mu m$, as used in Fig. 7 ($\propto \lambda^{6/5}$, see also atmospheric effects in app. 8.5).

For the typical spectral resolution $\Delta \nu_S$ we use different values motivated from HD and DD, from which one value, $\Delta \nu_S = 3$ GHz, is used for both and therefore the respective curves (in blue in Figs. 4-7) can be directly compared by the reader visually. On the one side, it is the highest resolution yet demonstrated by the VLTI-AMBER instrument in the K-band and the VLTI-MIDI instruments in the M-band (around 5 μm). On the other side, it is for HD a typical total bandwidth of current receivers, and could be a typical channel width of future receivers with 50-100 GHz total band width. Additional values are used to that: For DD, $\Delta\nu_S = 3$ THz stands for the lowest resolution ($R \approx 50$) possible in the K-band. For HD we use additionally $\Delta\nu_S = 100$ GHz for the total bandwidth of envisioned ultra-broadband HD receivers and 100 MHz as a typical submm-HD channel width of current receivers, e.g. in the case of ALMA. The same channel widths are also taken for two-telescope interference.

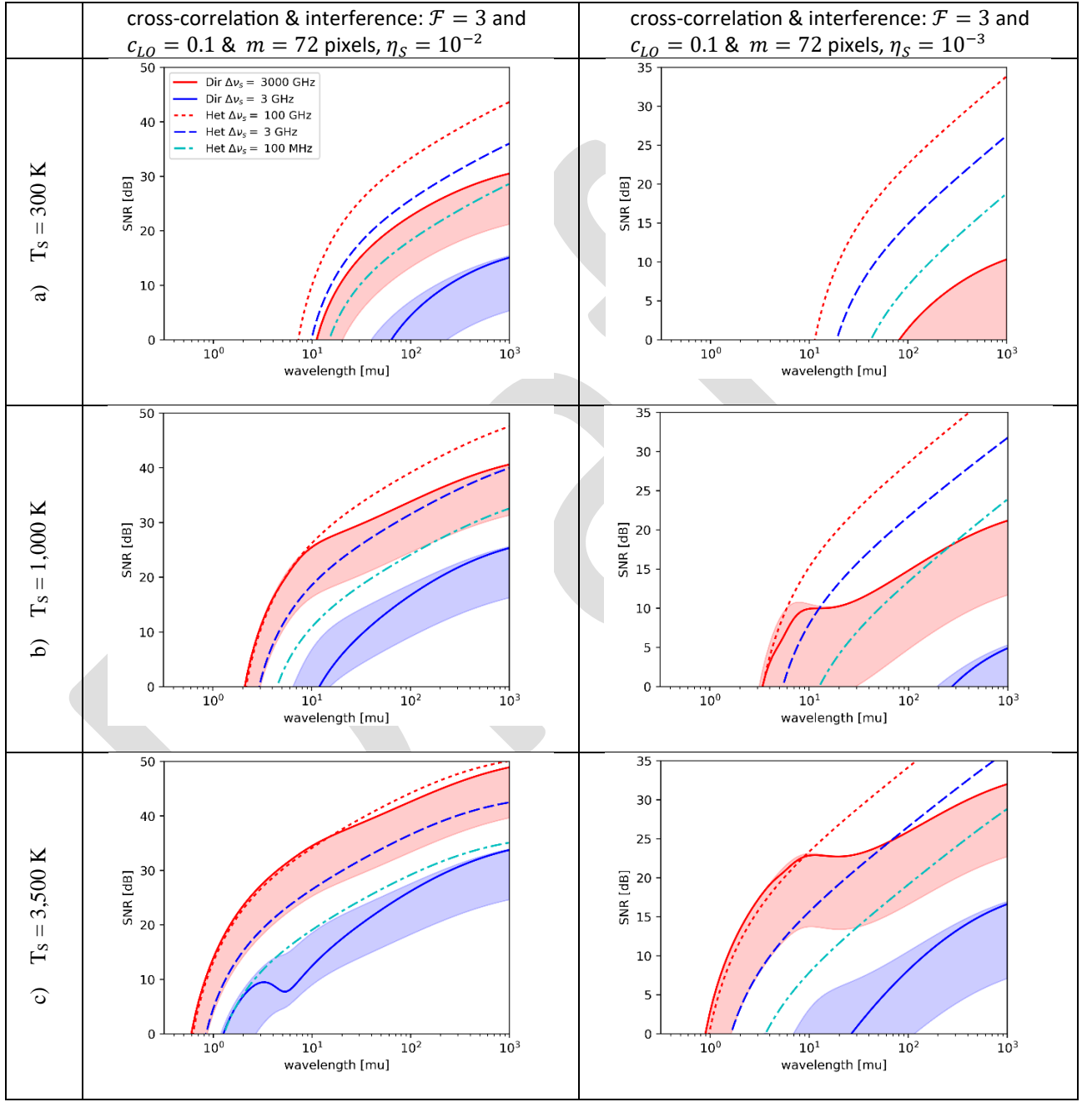

*Fig. 5: Two-telescope interferometry post-detection SNR comparison over wavelength with the parameters $T_B = 300$ K, $\Delta M_A = 1$, $\mathcal{T} = 0.9$, and $\Delta t = 10$ ms: HD (eqn. (45), for $c_{LO} = 0.1$, channel bandwidths $\Delta \nu_S = 100$ MHz, 3 GHz and 100 GHz) versus DD (eqns. (46)+(49), for $m_{pixel} = 72$, $\Delta \nu_S = 3$ GHz and 3 THz, shaded are the limits explained in Fig. 3), for $\eta_S = 10^{-2}$ (left) and $\eta_S = 10^{-3}$ (right), $c_S = 0.1$, and astronomical source temperature a) $T_S = 300$ K, b) 1,000 K, and c) 3,500 K. $N_{read}$ is assumed with a transition from 0 to $10^5$ in the range of 4-6 $\mu m$, and else constant. The blue curves can be compared directly. Included is also the phase deteriation factor of eq. (A46).*

As typical astronomical source temperatures we use a) $T_S$ = 300 K for warm dust in planetary accretion disks, b) 1,000 K for young hot planets, and c) 3,500 K for red giants or dwarfs. For single-telescope observation the beam filling factor is set to $\eta_S = 1$ and $\eta_S = 0.01$, while for two-telescope interferometers it is set to 0.01 and 0.001.

Please note that due to single-mode observation, the resulting post-detection SNR-plots are dominated in their spectral functional behavior (Fig. 4 & 5) by the mode occupation number of the source radiation field, $\bar{n}_S(\nu)$, and not by its Planck spectrum $B_{S,\nu}/h\nu \propto \nu^2 \bar{n}_S(\nu)$. Therefore, coming from long wavelengths, the SNR drops to 1 around the maximum of $B_{S,\nu}$.

### 3.4.1. Comparison of SNRs for single telescopes

In Fig. 4 (spectral, absolute) and in the left columns of Figs. 6 (over power) and 7 (spectral ratio) we compare the exact expressions for the post-detection SNRs for single telescopes (single-pixel detectors).

For the plots in Fig. 6, we get from eqns. (36) and (46):

$$\frac{SNR_{HD}}{SNR_{DD}} = \frac{SNR^1_{HD}}{SNR^1_{DD}} = \qquad (50)$$

$$\frac{\mathcal{T}\eta_S\bar{n}_S(\mathcal{T}\eta_S\bar{n}_S + 1) + \Delta M_A (1-\mathcal{T})\bar{n}_B\big((1-\mathcal{T})\bar{n}_B + 1\big)}{\mathcal{T}\eta_S\bar{n}_S(\mathcal{T}\eta_S\bar{n}_S + 1 + (1-\mathcal{T})\bar{n}_B + 1)}$$

For very high transmissions, $\mathcal{T} \to 1$, implying very low thermal background, we get

$$\frac{SNR_{HD}}{SNR_{DD}} \to \frac{\eta_S\bar{n}_S + 1}{\eta_S\bar{n}_S + 2} \qquad (51a)$$

which means that at high filling factors and high source occupation numbers (higher source temperatures and longer wavelengths) HD is equally sensitive as DD, while at short wavelengths it is a factor of two less.

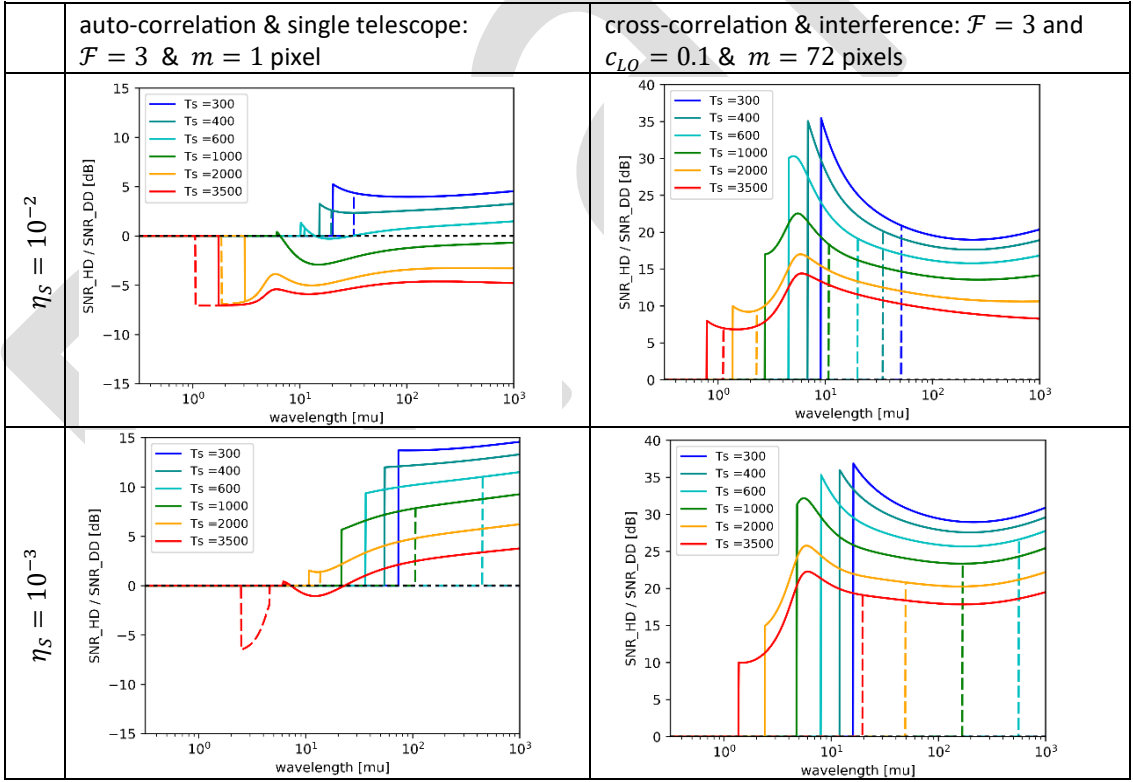

*Fig. 6: Expected ratio of the AC- (left) and CC- (right) post-detection HD-SNRs over DD-SNRs at different source temperatures $T_S$ for $\mathcal{T} = 0.9$, $T_B = 300$ K for same bandwidth and integration time, for $\eta_S = 10^{-2}$ (upper) and for $\eta_S = 10^{-3}$ (lower), and $c_S = 0.1$. The full-line cutoffs indicate the wavelengths where the HD-SNRs become unity, and the dashed-line cutoffs where the DD-SNRs become unity, both for $\Delta\nu_S = 3$ GHz and $\Delta t = 10$ ms, due to the dropping mode occupation number of the source (compare with Fig. 4 and 5). At higher bandwidth integration-time products those shift to shorter wavelengths. The phase deteriation factor of eq. (A46) is not included here, but would give +3 dB at the DD-SNR-cutoffs and an even more diverging ratio between the shown DD- and HD-cutoffs.*

For any transmission $\mathcal{T} < 1$, implying a thermal background which is mostly strong against the astronomical signal, the situation is more complicated. To illustrate it, we first consider an approximation, but which is not assumed for the plots: If we restrict our comparison to wavelengths shorter than 50 μm, to filling factors interesting for interferometry ($\eta_S \leq 0.01$) and to source temperatures not higher than 3500 K, we can assume $\mathcal{T}\eta_S\bar{n}_S < 0.02$ and with this the approximation for AC:

$$\frac{SNR'_{AC,HD}}{SNR'_{AC,DD}}$$
$$\approx \frac{\mathcal{T}\eta_S\bar{n}_S + (1-\mathcal{T})\bar{n}_B\big((1-\mathcal{T})\bar{n}_B + 1\big)}{\mathcal{T}\eta_S\bar{n}_S\big((1-\mathcal{T})\bar{n}_B + 2\big)} \quad (51b)$$

This ratio is $> 1$ if

$$\frac{\bar{n}_B}{\bar{n}_S} > \frac{\mathcal{T}\eta_S}{1-\mathcal{T}} \quad (51c)$$

Towards shorter wavelengths the left ratio drops below the right constant and so DD becomes better. In some cases, $SNR'_{DD} = 1$ or $SNR'_{HD} = 1$ is reached before that happens. In Fig. 6, left side, these cases are denoted as full (HD) or dashed (DD) cutoffs. For source temperatures smaller than the background temperature eq. (51c) is always fulfilled.

In eq. (51b) we can see for $\mathcal{T} < 1$ towards smallest filling factors, $\eta_S \to 0$, that the ratio diverges in favor of HD. The reason for this is the quadratic signal power dependence of DD at source powers small against the thermal background radiation power, while HD stays linear there in function of signal power $P_S$ (in the quadratic definition of the pre-detection SNR). This can be seen well in Fig. 7.

### 3.4.2. Comparison of the interference SNRs

In Fig. 5 we compare the HD correlation with the DD interference SNRs for the same parameters as in Fig. 4.

The filling factor of $\eta_S = 0.01$ was taken over from the single-telescope plots as an example for smaller relative baselines (e.g. VLTI with the UTs). As an even smaller

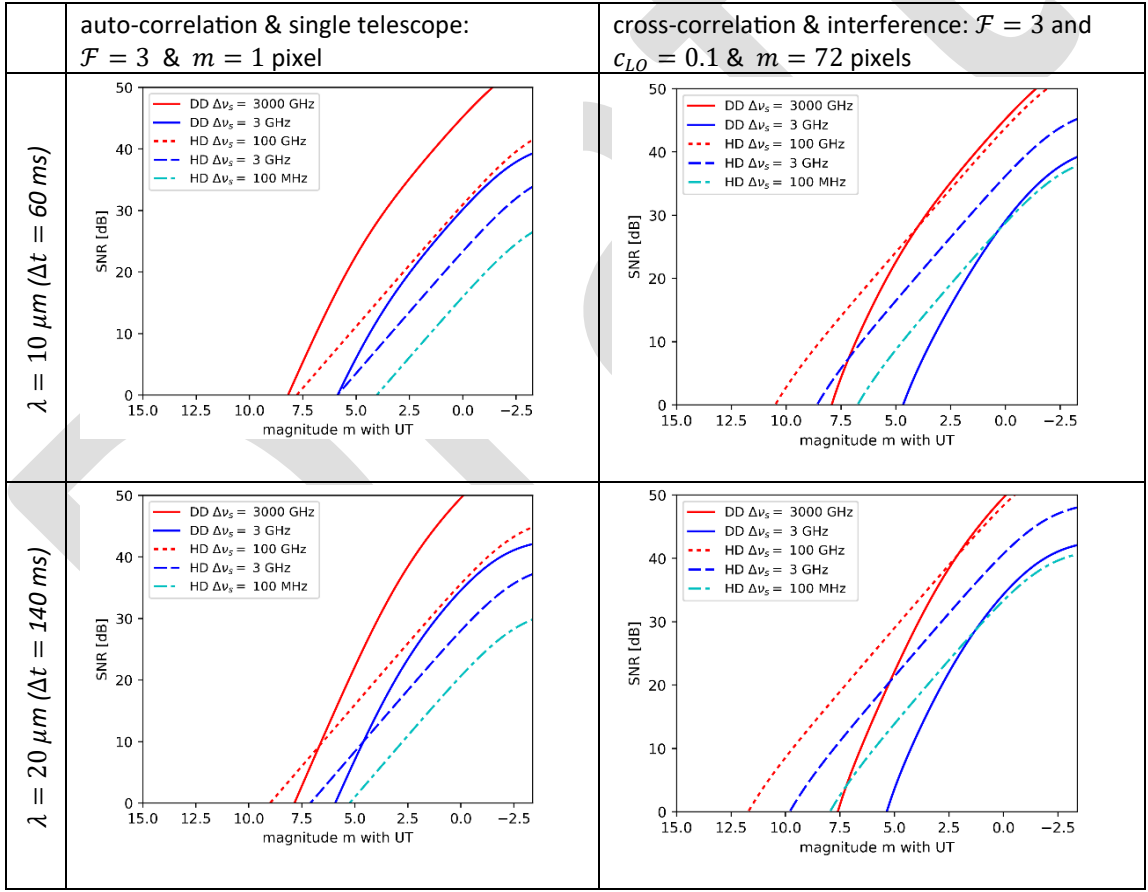

*Fig. 7: Post-detection SNR-comparison as a function of the coupled source power in a range from about $\eta_S = 10^{-7}$ to 1, using the relation $m_{AB} = -2.5\log_{10}(h\nu\eta_S\bar{n}_S/(A_T \cdot 3631 \cdot 10^{-26}))$, for the example of 8.2m-telescopes. With the blue curves, HD can be compared directly against DD (same bandwidth). The respective integration time is the indicated coherence time for the wavelength. $T_B = 300$ K, $\mathcal{T} = 0.9$, $\eta_Q = 0.7$, and plots are the same for all $T_S$, only the $\eta_S$-range shifts. As in the other figs., for HD the AC system noise temperature is assumed an experimental factor of $\mathcal{F} = 3$ above the quantum limit. Left (single-telescope/auto-correlation): DD is considered with one pixel ($m = 1$). Right (2-telescope-interference/cross-correlation): HD is considered with the improving factor $c_S = 0.1$ and DD is considered with the deteriorating factor of eqn. (49) with $m = 72$ and $N_{read} = 10^5$. On the right side, also the phase deteriation factor of eq. (A46) is included, assuming no fringe-tracking.*

filling factor, $\eta_S = 0.001$ was assumed (e.g. VLTI with the ATs).

Generally, the beam filling factors which appear in imaging interferometry are very small and of the order $\eta_S = (1.22 D/b)^2$, with $D$ the telescope diameter and $b$ the mean baseline of the array, or alternatively, with the Gaussian beam divergence and waist, $\eta_S = (\Theta_I/\Theta)^2 = ((\lambda/b)/(\lambda/\pi w_0))^2 = (\pi w_0/b)^2$. For example, ALMA has $\eta_S = 2 \cdot 10^{-6}$ at 10km-baselines, VLTI has just $\eta_S = 1 \cdot 10^{-2}$ at 100m-baselines using the UTs, but $\eta_S = 5 \cdot 10^{-4}$ using the ATs. The highest spatial resolution interferometry at extreme baselines, as envisioned with future interferometry projects, results in mid-IR filling factors comparable to that of ALMAs largest baselines. Finally, with 8m-telescopes at baselines up to 2km, as projected for the Planet Formation Imager [54][55], objects have to be detected in one of the 250 two-telescope sub-lobes within the single-telescope main-beam. The smallest object still not resolved has then a single-telescope filling factor of the order of $10^{-5}$. Note that independently of this, the number of fringes within the PSF in the interferometric focal plane on the detector array is usually made to be much less than the interference-lobe number on-sky as appearing in eqn. (48), depending solely on the collimated telescope ("relay") beams combination angle on the detector, see different notation in eqn. (A37) of the appendix 8.4.

The interferometry SNR is basically the single telescope SNR multiplied by the visibility, and for DD with a deteriorating factor between 1 and $1/\sqrt{m} < 1$ according to eqn. (49) and Fig. 3, whereas for HD with the improving factor of $\times 1/c_{LO} > 1$ according to eqn. (45), if balanced mixers are used.

$$\frac{SNR^1_{CC,HD}}{SNR^1_{V,DD}} \approx \frac{SNR^1_{AC,HD}}{SNR^1_{DD}} \cdot R_{int} \qquad (52a)$$

where

$$R_{int} \coloneqq \frac{2}{c_{LO}} \times \sqrt{\frac{N_{rad} + mN_{read}}{N_{rad} + N_{read}}} \qquad (52b)$$

which, for a BCR-system for HD and 72 pixels for DD, would have a value in the range $20 \cdots 170$, depending on the wavelength, channel width, integration time and DD read/dark counts.

To be plotted in Fig. 6, eqns. (50) and (52a) are further multiplied by the ratio of the post-detection NR-factor of $(1 + NR^1_{HD})/(1 + NR^1_{DD})$, which is the same for AC and CC, giving an increase in favor of HD towards shorter wavelengths, so that the ratios are peaking towards the cutoffs. The latter indicate the additional condition of $SNR_{CC,HD,post} > 1$ for $\Delta \nu_S = 3$ GHz and $\Delta t = 10$ ms, and can be also inferred from Figs. 5 and 6. At the small beam filling factors prevalent in interferometry, the advantage of HD is impressively high for interferometry, a point which was not indicated before in literature.

In Fig. 7 the SNRs are plotted over the astronomical magnitude observed with a 8.2 m-telescope (e.g. the UTs at the VLTI), see eq. (3a'). For any other telescope, e.g. for the smaller AT-telescopes (1.8 m) of the VLTI, it holds $m_{AB,UT} - m_{AB,AT} = 2.5 \cdot \log_{10}(A_{UT}/A_{AT})$. The plots are over the beam filling factor $\eta_S$ and the source occupation number $\bar{n}_S$ both simultaneously (in product). Therefore, they are given independently of $T_S$, as only the $\eta_S$-range would shift. The magnitude-definition is assumed here wavelength-independent because only relative brightness is needed for comparison between HD and DD.

From all the plots we see that for interferometry at wavelengths larger $\lambda = 5$ μm, future ultra-broadband-HD is consistently more sensitive for ground-based observations in the frequent cases of a transmission significantly less than unity in narrow atmospheric windows (e.g. the Q-band). However, for a space-born situation there should not be a significant difference from that situation: At the higher sensitivities achievable in space with envisioned future lower readout-noise DD detector-arrays, the lower zodiacal light background needs to be considered analogously to the higher atmospheric background at the lower sensitivities achieved from the ground with current detectors.

## 4. Discussion

We want to draw the reader's attention to several important, previously controversial, points resolved in or for this paper:

1. Due to its fundamental character, our experimental result of a 20-fold enhanced SNR in cross-correlation of two balanced receivers compared to a single balanced receiver [17] needed more confirmation. We deemed it necessary because a non-thermal SLED test source was used for first (because of higher intensity) and since each receiver is separately operating within a factor of 4 close to the noise temperature quantum limit, this infers a cross-correlation sensitivity well below the quantum-limit. With a bandwidth of 7 THz compared to the SSB receiver channel-width of 3 MHz, the SLED could be regarded as incoherent enough. However, the remaining doubt was that it could still have quantum properties, e.g. less photon bunching, which would make it behave rather coherent compared to a true thermal source. However, this concern was resolved recently experimentally, as we could reproduce and refine the previously reported result at $\lambda = 1550$ nm with a true thermal source [18]. Furthermore, the improved noise behavior in cross-correlation, predicted to be wavelength-independent [17], could be observed thereafter also in the sub-millimeter-range using two balanced receivers at 460 GHz [18]. Also, a proof with full quantum theory was proposed by others in the meantime [56].

2. The consequent incorporation of both, the beam-filling factor $\eta_S \ll 1$ (interferometry) and the atmospheric transmission $\mathcal{T}$ and emission $(1 - \mathcal{T})$ into the derivations of both, the HD and DD SNRs for single and two-telescope detection through $P_S = P_{S,\nu} \Delta \nu = \mathcal{T} \eta_S h\nu \bar{n}_S \Delta \nu$ and $\overline{\delta n_S^2} = \eta_S \mathcal{T} \bar{n}_S (\eta_S \mathcal{T} \bar{n}_S + 1)$ seems to not have been formulated out in previous publications. With the motivation to consider the effects of interferometry and atmosphere, we had to introduce the notation of modes with occupation numbers.

3. For HD we corrected the misconception of others (e.g. [49]) assuming in post-detection the cross-correlation power to be of square-law dependence on the source power: Rather, it must be of the same form as the auto-correlation power, having a linear dependence on the source power.

4. In consequence of the linearity of CC, it can be arrived at the usefulness of the definition of a (different) noise temperature (in dependence of the parameter of the phase difference) for cross-correlation in analogy to auto-correlation[17], which obviously has not been considered previously.

5. There is the common believe that HD gets worse towards the visible whereas DD not, because the quantum noise is rising proportionally with the frequency, $T_Q = h\nu/k_B$, as manifested also in eqn. (37) by dividing through the factor $h\nu$. We see from eqn. (47a) that the same factor occurs also in DD. Here also, quantum noise increases with frequency, but here not that of the signal amplified by the strong local oscillator but that of the weak signal alone. This has the same effect on the final SNR-formula.

6. For DD it was derived quantitatively the deterioration of the post-detection SNR by the distribution of the signal onto many pixels. The pixel number necessary is in the all-into-one detector-array beam-combination scheme proportional to the baseline number and therefore to the telescope number squared. For the PFI (projected 20 telescopes) we expect 5 times more pixels, i.e. 360.

7. We introduced for the comparison of HD vs DD, that all technicalities of post-detection noise-contributions are condensed into the "post-detection noise-ratio", $NR$. This number is comparable for both detection schemes under most conditions. However, for DD, $NR$ diverges towards low photon numbers, i.e. towards the detection limit and/or towards visible wavelengths.

8. In the presence of thermal background, the roll-off of the DD SNR towards weakest signals is of square law behavior while that of HD is linear. This makes HD more sensitive than DD towards disappearing source power and renders HD advantageous for lower atmospheric transmission values, as are for example common in the Q-band (20-40 µm).

9. In consequence of the above points, a) regarding a single baseline, at $\lambda = 10$ µm a DD-interferometry-system with 3 THz channel-width (R=10, lowest possible resolution in DD) is in terms of source faintness at SNR=10 equivalent to a HD-system with 3 GHz channel-width, and at $\lambda = 20$ µm even to a HD-system with 0.3 GHz channel-width (see Fig. 7, right side). A BCR-HD-interferometry system has a detection limit fainter than a 72-pixel DD-interferometry system by 4 magnitudes (a factor of 40) compared at the same channel-width, and it reaches fainter at SNR=10 about 3 magnitudes (a factor of 16). b) This sensitivity-advantage of HD should increase with the number of baselines, since for DD the pixel number $m$ must increase in the all-in-one detector combination scheme, while in HD the use of bi-spectrum real-time combination before integration is rather expected to reduce the noise of the closure phase and therefore increase the sensitivity of the total system. c) Having shown that the effect of 1.) is independent of the wavelength, HD should have a lower detection limit than DD for all thermal infrared wavelengths. It means that any existing cross-over of HD to DD should be shifted towards higher frequencies, relative to that of single-mixer receivers for which it was calculated to be in the few-THz-range [23].

## 5. Conclusion

We derived from first principles of a semi-quantum theory single-baseline interferometric SNR formulas for DD and HD in a common formalism, to render a full comparison in spectral and source power dependence considering recent developments of system architectures and atmospheric emission/absorption. The results encourage to develop super-broad-band mid-infrared heterodyne technologies for future interferometry facilities like the currently discussed Planet Formation Imager (PFI) [54][55], the extension of current IR-interferometry systems [21], or the application to receiver systems in the submm- and far-infrared range.

## 6. Acknowledgements

During the development of the presented results E.A.M. acknowledges support from Fondecyt Regular 1201722, ALMA 31190064, ANID, Chile, and from SFB 956, DFG, Germany, and fruitful discussions with L. Labadie and C.E. Honingh. He also acknowledges previous support from CONICYT project. This work was realized on basis of the long-term framework of the previous CONICYT Chile grants ALMA 31080020, 31090018, and 31110014, 31140025, Quimal 150010 and Fondecyt Regular 1090306, and Basal AFB-170002, with additional support from the ESO-Mixed Committee fund Chile assigned in 2016 and in 2018, including support for M.H. during 2019-20. For after this time, M.H. acknowledges support by the Chinese Academy of Sciences (CAS), through a grant to the CAS South America Center for Astronomy (CASSACA) in Santiago, Chile. E.M. acknowledges his Fondecyt Postdoctorado 3180130.

# 8. Apendices

## 8.1. Considerations between radio and optical detection under thermal background

In this appendix we gathered side-conclusions from the fundamental relations in chapter 2.

### 8.1.1. The different conventions in noise characterization

As for point-like astronomical sources the beam filling factor is $\eta_S \ll 1$, and because it is hard for direct detection in ground-based systems to baffle the telescope étendue down to $\Delta M_A = 1$ and to have a low effective background (sky and optics) radiation temperature, we usually have $|\delta P_B| \gg |\delta P_S|$ in a ground-limited situation. This implies a source intensity-independent noise contribution so that the background-limited noise-equivalent power ($NEP$) is defined as (with both polarizations):

$$\delta P_B = 2h\nu \sqrt{\Delta M_A \, \bar{n}_B(\bar{n}_B + 1)\Delta \nu \Delta f} =: NEP\sqrt{\Delta f} \quad (A1)$$

This is first discussed independently of the detection method for looking at an optically thick thermal foreground ($\mathcal{T} = 0$). Reduced to a single polarization, $\delta P_B' = \delta P_B/\sqrt{2}$, the radiation-$NEP$ of eqn. (A1) can be related to the noise-equivalent temperature variations due to the background, which a hypothetical receiver would see in each mode, and such we could define as:

$$\delta P_B' =: \Delta M_A \, k_B \delta T_B \, \Delta \nu_S \quad (A2)$$

For $\bar{n}_{B/S} \gg 1$ (long-wavelengths, high temperatures) we have from eqn. (11), looking at an optically thick medium ($\mathcal{T} = 0$):

$$\delta P_B' = h\nu \sqrt{\bar{n}_B(\bar{n}_B + 1) \cdot \Delta \nu_S \, 2\Delta f} \quad (A3a)$$

$$\approx h\nu \bar{n}_B \Delta \nu_S \sqrt{\frac{2\Delta f}{\Delta \nu_S}} = P_B/\sqrt{\Delta \nu_S \Delta t}$$

To make now a relation to the system noise temperature $T_{rec}$ of a coherent receiver, we consider the following: The Y-factor in a receiver noise-temperature "hot-cold" measurement is defined as $Y = P_{out,hot}/P_{out,cold}$ (see e.g. [17]). Using as the hot input temperature in such a measurement exactly one which doubles the output noise power (the special value Y=2), and as cold input temperature $T_B = 0$, we get $T_{rec} \propto P_{S,hot}$ and $\delta T_{rec} \propto \delta P_{S,hot}$, and eqn. (A3a) converts to

$$\delta T_{rec} = T_{rec}/\sqrt{\Delta \nu_S \Delta t} \quad (A3b)$$

which is the well-known "radiometer formula/equation". In its alternative derivation in Wilson et al. [39], the temporal mode number, eqn. (1c), appears as the number of independent measurements of the Nyquist sampling theorem.

Firstly, we can obtain from this the relation between (two-polarization) $NEP$ and (single-polarization) $T_{sys}$ for the long-wavelength range:

$$\delta P_B' = \Delta M_A \, k_B T_{sys} \sqrt{2\Delta \nu_S \Delta f}$$

$$\Rightarrow \quad NEP \approx 2\Delta M_A \sqrt{\Delta \nu_S} \, k_B T_{sys} \quad (A4)$$

Conventionally, the direct detection community uses the noise-equivalent power (NEP, unit $W/\sqrt{Hz}$) as the sensitivity measure, while the heterodyne community uses the system noise temperature (noise-equivalent spectral power density, $NESPD = k_B T_{rec}$, unit $W/Hz$). One can see that the different units mainly prevail from the very different wavelength ranges in which the definitions were first made. Note that $T_{sys}$ has a single-mode meaning while $NEP$ has in general a multi-mode meaning. Therefore, to compare both detection schemes, it is best to express both sensitivities in dimensionless signal-to-noise ratios ($SNR$).

### 8.1.2. The traditional SNR-formula in the radio range

Secondly, we can readily obtain for this case ($\bar{n}_S \gg 1$) from eqn. (A3b), replacing the source for the background, the (unsquared) SNR-formula of a long-wavelength receiver (known from radio-astronomy)

$$SNR_{radio,HD} = \frac{\Delta P}{\delta P} = \frac{\Delta T_A}{\delta T_{rec}} = \frac{\Delta T_A}{T_{rec}} \sqrt{\Delta \nu_S \Delta t}$$

$$\leq \frac{\Delta T_A}{T_{rec,QL}} \sqrt{\Delta \nu_S \Delta t} \quad (A5a)$$

$$= \frac{\bar{P}_{\nu,S}}{h\nu} \sqrt{\Delta \nu_S \Delta t}$$

where $\Delta T_A = \bar{P}_{\nu,S}/k_B$ is the antenna temperature increase due to the source, and $T_{rec,QL} = h\nu/k_B$ is the receiver noise temperature quantum limit ($T_{rec} \geq T_{rec,QL}$) in case of coherent receivers. This limit would be given by the laser shot noise (or equivalently by the ZPFs), but it is included here only phenomenologically while it is properly derived in the main part of the article. While the formula above is derived just considering source and background, the system noise temperature must be derived from the SNR considering also the local oscillator noise. More generally, it can be determined by setting $k_B T_{sys} \coloneqq NESPD = P_{\nu,S}/SNR(P_{\nu,S})$, ($NESPD$ = "noise-equivalent spectral power density") were the form of $SNR(P_{\nu,S})$ has to be (or made) linear in $P_{\nu,S}$, see eqn. (35) in [17].

Eqn. (A5a) should also be valid for an incoherent detector in the background-limited case, in which we would replace $T_{rec}$ by $T_B$. Here it seems that $T_{rec} \propto \nu_S$ while $T_B = const$, so that DD would become overwhelmingly better towards the visible. Coming directly from eqn. (A3a) it is at long wavelengths

$$SNR_{radio,DD} = \frac{P_S}{\delta P_B'} \approx \frac{P_S}{P_B} \sqrt{\Delta \nu_S \Delta t} = \frac{P_{S,\nu}}{h\nu \, n_B} \sqrt{\Delta \nu_S \Delta t}$$

while it is at shorter wavelengths:

$$SNR_{IR-vis,DD} \approx \frac{P_{S,\nu}}{h\nu \sqrt{n_B}} \sqrt{\Delta \nu_S \Delta t} \quad (A5b)$$

identical with the expression used in [29] and [30]. However, not considered there is the dominating radiation noise of the signal itself when $n_B \to 0$. Therefore, these expressions, considered in [28], [29], and [30], generating there a huge advantage of DD over HD, are wrong towards shorter wavelengths and/or high transmissions, already starting in the mid-infrared, where then DD becomes comparable to HD, see section 3. Furthermore, it is to be noticed that these are pre-detection SNRs.

### 8.1.3. Equivalent SNR-formula in the visible range

According to the above we would get for this case ($\bar{n}_S \ll 1$) from eqn. (12) $\delta P_S' \approx h\nu \sqrt{\eta_S \bar{n}_S \Delta\nu_S\, 2\Delta f} = \sqrt{h\nu P_S}/\sqrt{\Delta t}$, and so an equivalent SNR for no background as:

$$SNR_{vis} = \frac{P_S}{\delta P_S} = \sqrt{\frac{\bar{P}_{v,S}}{h\nu}} \sqrt{\Delta\nu_S \Delta t} \qquad (A6)$$

This derivation of the SNR is just in the presence of the noise from the signal itself, without a thermal background. In case of HD, again this result is phenomenologically argued, and no laser noise is included.

### 8.1.4. HD pre-detection SNR in the literature

In the traditional literature deriving the SNR for HD, e.g. in [24] or [33], it is considered that the noise contributions of eqns. (31) are detected directly (like in DD) by the single-mixer detectors, omitting the contributions generated by the down-conversion. Then, those DD contributions are simply to be added and give for the total radiation noise,

$$\overline{|\delta P_{tot}|^2} = 2h\nu\Delta f \{(\mathcal{T}\eta_S \bar{n}_S + 1) \cdot P_S \qquad (A7)$$
$$+ ((1-\mathcal{T})\bar{n}_B + 1) \cdot P_B$$
$$+ F\, P_{LO}\}$$

It follows:

$$SNR_{het,I} := \frac{(P_{het})^2}{|\delta P_{tot}|^2} = \frac{P_S P_{LO}}{|\delta P_S|^2 + |\delta P_B|^2 + |\delta P_{LO}|^2} \qquad (A8)$$

$$= \frac{P_S/2h\nu\Delta f}{\frac{[(\mathcal{T}\eta_S \bar{n}_S + 1)P_S + ((1-\mathcal{T})\bar{n}_B + 1)P_B]}{P_{LO}} + F}$$

For $P_{LO} \gg P_B, P_S$ and a single temporal mode ($\Delta f = \Delta\nu/2$) it is:

$$SNR^1_{het,I} \approx \frac{P_{S,\nu}}{h\nu\, F} = \frac{\mathcal{T}\eta_S \bar{n}_S}{F} \qquad (A9)$$

This expression is determined by the excess laser noise $F$. Entered into eqn. (24), omitting the NR-denominator, it gives the traditional SNR-formula which is usually cited in the literature, e.g. [9], [22], [23], [24], [25], [26], [27], [44], [33], [28], [30], mostly citing [31], but where the derivation is not really apparent.

$$SNR_{het,I,post} \approx \frac{P_{S,\nu}}{h\nu\, F}\sqrt{\Delta\nu \Delta t} = \frac{P_S}{h\nu\, F}\sqrt{\Delta t/\Delta\nu} \qquad (A9b)$$

In the references [28], [29], and [30] it was assumed that the HD NEP has to be $NEP_{het} = h\nu\sqrt{\Delta\nu/\Delta t}$, directly given by the ZPF, in order to get expression (A9b). However, including the beam-filling factor and the atmosphere, we show that the expressions for the SNR cannot be as simple as there.

The corresponding expression for CC-SNR is therefore in analogy to eqn. (A8):

$$SNR'_{CC} := \left(\frac{S_{CC}}{N_{CC}}\right)^2 \qquad (A10)$$

$$= \frac{\gamma P_S P_{LO}}{c_S|\delta P_S|^2 + c_B|\delta P_B|^2 + c_{LO}|\delta P_{LO}|^2}$$

$$= \gamma P_S P_{LO}/\{c_S 2h\nu\Delta f\, (\mathcal{T}\eta_S \bar{n}_S + 1) \cdot P_S$$
$$+ c_B 2h\nu\Delta f((1-\mathcal{T})\bar{n}_B + 1) \cdot P_B$$
$$+ c_{LO} 2h\nu\Delta f\, F\, P_{LO}\}$$

$$= \frac{\mathcal{T}\gamma\eta_S \bar{n}_S\, (\Delta\nu_S/2\Delta f)}{\frac{[c_S \mathcal{T}\eta_S \bar{n}_S(\mathcal{T}\eta_S \bar{n}_S + 1) + c_B(1-\mathcal{T})\bar{n}_B((1-\mathcal{T})\bar{n}_B + 1)]\Delta\nu_S}{\bar{n}_{LO}\,\Delta\nu_{LO}} + c_{LO}F}$$

and for $\delta P_{LO} \gg \delta P_S$ and $P_{LO} \gg P_S$ this gives for a single temporal mode:

$$SNR_{CC,I} \approx \frac{\gamma}{c_{LO}} \frac{\mathcal{T}\eta_S \bar{n}_S}{F} = \frac{\gamma}{c_{LO}} \cdot SNR_{het,I} \qquad (A11)$$

Here, no assumption about $c_S$ is necessary, since it does not appear in the final formula. With a single beam splitter and no balanced photodiodes we obtain obviously in this theory that the SNR is not improving in CC, since with a single power splitter for the LO we have $c_{LO} = 1$.

### 8.2. Derivation of the noise ratio factor NR

To derive the form of the noise ratio factor

$$NR^1 := \frac{N^1_{el}}{Z\Re^2 \overline{(\delta P^1_{rad})^2}}$$

defined in eqn. (23), we first determine the full form of $N^1_{el}$. The upper index 1 refers the single temporal mode time scale (before integration). We model the quantum efficiency with a corresponding attenuation in front of the detector and use quantum noise propagation, eq. (9), $\overline{\delta n'^2} = \eta_Q(1-\eta_Q)\bar{n} + \eta_Q^2 \cdot \overline{\delta n^2}$. After the attenuation (denoted by the prime ′) we have according to eq. (7):

$$\overline{\delta I_{el,rad}^2}/\left(\frac{e}{h\nu}\right)^2 = \overline{(\delta P^1_{rad}')^2} = 2(h\nu)^2 \Delta\nu \Delta f\, \overline{\delta n'^2}$$

and with eq. (9) and $P_{rad} = h\nu\, \bar{n}\, \Delta\nu$ we obtain:

$$\overline{\delta I_{el,rad}^2} = \Re^2 \left[\overline{\delta P_{rad}^2} + 2h\nu P_{rad}\Delta f\frac{(1-\eta)}{\eta}\right] \qquad (A12)$$
$$= \Re^2 \overline{\delta P_{rad}^2} + (1-\eta)\, 2eI_{ph}\Delta f$$
$$=: \overline{\delta I^2}_{rad,converted} + \overline{\delta I^2}_{sh,ph}$$

The second term, the shot noise variations of the photocurrent in the detector, is statistically independent of the first term, the exactly converted radiation noise. Therefore, we count it from now on to the post-detection electronic noise, which is also statistically independent from the radiation noise. This did not appear yet in that form in the literature, see e.g. [31]. It means that at high quantum-efficiencies there is less uncorrelated shot noise current contained in the photocurrent and the electron fluctuations are following the radiation fluctuations with high correlation. This was characterized already in some experiments, see e.g. [43], [57], and [58].

This taken together is projected to the detector output:

$$N_{el}^1 = Z\overline{\delta I^2}_{sh,ph} + Z\overline{\delta I^2}_{sh,dark} + Z\overline{\delta I^2}_{therm}$$
$$= \left[Z2e\left((1-\eta_Q)I_{ph} + I_{dark}\right) + k_B T_{ampl}\right]\Delta f^1 \quad \text{(A13)}$$

$\Delta f^1 = \Delta\nu/2$ is the fluctuation bandwidth for a single temporal mode before integration. For DD there is no amplification before integration (see fig. 1), therefore the term $k_B T_{ampl}$ is omitted. Instead, readout noise together with a distribution of the light to parallel pixels leads to a degradation of the post-detection-SNR, as derived in appendix 8.4. Next, we determine the denominator of eq. (23).

### 8.2.1. NR for HD

For HD (single spatial mode and single polarization) the converted radiation noise power is for single mixers ($F > 1$):

$$Z\overline{\left(\delta I_{el,rad}^1\right)^2} = Z\mathfrak{R}^2\overline{(\delta P_{rad}^1)^2}$$
$$\approx Z\frac{\eta_Q e}{h\nu}[2Fh\nu(\mathfrak{R}P_{rad})\Delta f^1] \quad \text{(A14a)}$$
$$= F\eta_Q Z\, 2eI_{ph}\,\Delta f^1$$

so that

$$NR_{het}^1 = \frac{\left[Z2e\left((1-\eta_Q)I_{ph} + I_{dark}\right) + k_B T_{ampl}\right]}{F\eta_Q Z\, 2eI_{ph}}$$
$$= \frac{1}{F}\left[\frac{(1-\eta_Q)}{\eta_Q} \right. \quad \text{(A15a)}$$
$$\left. + \frac{Z2eI_{dark} + k_B T_{ampl}}{\eta_Q Z\, 2eI_{ph}}\right]$$

The second term tends towards zero, if the LO-power can be made large enough and $I_{dark} > 0$ is tolerable ($I_{ph} \gg I_{dark}$), so that for higher quantum efficiencies, HD is almost ideal.

For balanced mixers, using $\overline{\delta P_{rad}^2} = |\delta P_{het,BPD}|^2 \approx h\nu(\mathcal{T}\eta_S\bar{n}_S + 1 + (1-\mathcal{T})\bar{n}_B + 1)P_{LO}2\Delta f$ from eqn. (35), the factor $F$ must be replaced by the factor $(\mathcal{T}\eta_S\bar{n}_S + 1 + (1-\mathcal{T})\bar{n}_B + 1)$. Then

$$NR_{het}^1 \approx \frac{1}{(1-\mathcal{T})\bar{n}_B + 2}\left(\frac{1}{\eta_Q}-1\right) \lesssim 0.5 \quad \text{(A16a)}$$

for $\eta_Q > 0.5$.

### 8.2.2. NR for DD

Using eqns. (7), (11) and (12), for DD the converted radiation noise power towards weak signals is for one polarization and a single temporal mode:

$$Z\overline{\delta I^2}_{el,rad} \approx Z\mathfrak{R}^2(h\nu)^2\big[\eta_S\mathcal{T}\bar{n}_S(\eta_S\mathcal{T}\bar{n}_S + 1)$$
$$+ (1-\mathcal{T})\bar{n}_B((1-\mathcal{T})\bar{n}_B + 1)\big]\Delta\nu_S\,\Delta f^1$$
$$\approx ((1-\mathcal{T})\bar{n}_B + 1)\,\Delta M_A\,\eta_Q Z\, 2eI_{ph}\,\Delta f^1 \quad \text{(A14b)}$$

for signals weaker than the background, so that

$$NR_{DD}^1 = \frac{\left[Z2e\left((1-\eta_Q)I_{ph} + I_{dark}\right)\right]}{((1-\mathcal{T})\bar{n}_B + 1)\,\Delta M_A\,\eta_Q Z\, 2eI_{ph}} \quad \text{(A15b)}$$

$$= \frac{1}{((1-\mathcal{T})\bar{n}_B + 1)}\left(\frac{1}{\eta_Q}\left(1 + \frac{I_{dark}}{I_{ph}^1}\right) - 1\right) \quad \text{(A16b)}$$

$I_{dark}$ is *not* related to the readout noise of the detector, $N_{read}$, which comes in only after integration (see scheme in Fig. 1, and the derivation in appendix 8.4.), but rather to the thermal excitation of electrons into the conduction band. This thermal equilibrium can be assumed to occur at very short time scales compared to even the single-temporal-mode time (coherence time). Whatever its value is, in DD there is the fundamental problem that $I_{dark}/I_{ph}^1 = N_{dark}/N_{rad}^1$ could possibly diverge when the total radiation power goes towards zero. However, in practice this does not occur because of cryogenic cooling and the non-vanishing radiation background. However, reduction of $I_{dark}$ by cooling goes normally at the expense of increasing the readout noise (see Aquarius detector for MATISSE).

### 8.3. Clarification SNR in HD cross-correlation

The assumption $SNR_{CC} \propto (SNR_{AC})^2$ is not correct. We regard this first for the relation $S_{CC} = S_{AC1} \cdot S_{AC2}$. Using Gaussian error propagation,

$$(SNR_{CC})^2 = \frac{(S_{AC1} \cdot S_{AC2})^2}{\left(\delta(S_{AC1} \cdot S_{AC2})\right)^2}$$
$$= \frac{(S_{AC1} \cdot S_{AC2})^2}{S_{AC2}^2 \overline{(\delta S_{AC1})^2} + S_{AC1}^2 \overline{(\delta S_{AC2})^2}}$$
$$= \frac{SNR_{AC1}^2 \cdot SNR_{AC2}^2}{SNR_{AC1}^2 + SNR_{AC2}^2}$$

In case of $SNR_{AC1} \approx SNR_{AC2}(=:SNR_{AC})$ it is $SNR_{CC} \approx SNR_{AC}/\sqrt{2}$, a linear relation between both. The result would be similar if we assumed more correctly $S_{CC} = \sqrt{S_{AC1} \cdot S_{AC2}}$. Then we obtain $SNR_{CC} \approx SNR_{AC}$. In fact, the result we obtain experimentally for the CC-SNR, see [17], is for small signals indeed a linear behavior like in AC, according to eqns. (21). Therefore, the conclusion that HD-interferometry suffers towards weak signals a drawback of orders of magnitude in SNR compared to DD-interferometry, as assumed e.g. in [49], is totally wrong. Rather, including atmospheric absorption, it is the other way around; the DD-SNR drops off quadratically at vanishing source power.

### 8.4. Impact of the number of pixels in DD to the post-detection visibility SNR

In DD the pixel dilution effect has to be considered as well, certainly in interferometry where the interference pattern has to be sensed with a line of pixels for each wavelength, and eventually also at single telescopes in the case of the Airy-pattern of the star distributed over several pixels. The quasi-exact solution to this is derived in the following. The complex visibility ((visibility modulus and phase) can be measured in two equivalent ways, spatially (Fizeau-configuration) or temporally. 1.) The spatial method works with detecting an interference pattern on a detector array, keeping the phase difference between the two telescopes constant. Along the detector array's pixel line the relative phase between the two telescope beams is changing, thus the fringe pattern is formed. This means that the photon number in a certain integration time is divided up between $m$ detectors imaging the pattern. To achieve precision in the visibility measurement, it is best to fit as many fringes as possible, therefore $m$ must be a sizeable number (see e.g. the MATISSE-instrument with $m = 72$), also because all telescope beams are superimposed on one CCD and the baselines are separated by Fourier-analysis. Furthermore, the offset of the fringes allows to determine the variation of the relative phase between the telescopes.

Since the pattern is periodical in case of small bandwidths, the absolute phase-difference can only be determined for very large bandwidths, when the fringe intensity drops off rapidly from the center-fringe, so that this so-called "white-light fringe" can be identified clearly. A photonic realization of this principle for fringe trackers (emphasis on phase) consists in a splitting of the signal beam power into four parallel channels ending in four pixels/detectors with different but fixed phase delays in the channels at the same time (see e.g. the GRAVITY instrument). 2.) The temporal method consists in stepping the optical path difference between the two telescopes between at least four values and sensing the resulting time-varying interference intensity on a single detector. Here the integration time is divided up into $m$ sub-intervals. Both methods therefore result in a division of the available photon number by $m$, $\bar{N}_{1px} = \bar{N}/m$.

In the Fizeau-geometry the interference-fringes in the beam-combiner instrument form the angled identical fundamental-mode Gaussian beams (in general Airy-patterns in the focal planes which contain higher Gaussian modes). In the CCD-array-plane this is equivalent of having a higher-order single mode with as many nodes (zeros) as interference fringes are visible within the pixel projection of the mode, $\psi(x_i)$, where $x_i$ is the coordinate of the pixels and the mode shall be normalized to $\sum_{i=1}^{m}|\psi(x_i)|^2 = 1$. Therefore, the transition probability of a photon from that mode into the aperture $\Delta x_{px}$ of one pixel, $\eta_{S,i}$, is given by the overlap of that higher-order mode with the $i$th pixel, so that the overall detection probability in a pixel becomes $\eta_i := \eta_S \eta_Q \eta_{S,i}$ for the signal, and $\eta_i = \eta_Q \eta_{B,i}$ for the background. According to eqns. (9)-(11), the occupation number variance after the distribution to the pixels and after detection is $\overline{\delta n'_i{}^2} = \eta_i \bar{n} [1 + \eta_i \bar{n}]$. For the signal, the pixel overlap $\eta_{S,i}$ is varying according to $x_i$ as

$$\eta_{S,i} = \frac{1}{m}[1 + V_{pre} \cos(k_{fringe} x_i + \Delta\varphi)]|\psi(x_i)|^2 \quad (A17)$$

where $|\psi(x_i)|^2$ is the projection of the sky-sensitivity $I_A(\varphi, \vartheta)$ of eqn. (48) onto the pixel-array and $V_{pre}$ is the optical pre-detection visibility. The background is uncorrelated noise and so does not form interference fringes, so that $\eta_{B,i} = (1/m)|\psi(x_i)|^2$. The counted electrons from each pixel are with this expressed as

$$N_{1px,i} = (\eta_{S,i} R_S + \eta_B R_B)\Delta t + \bar{N}_0 \quad (A18a)$$

with the photon rates according to eqn. (3) as

$$R_S = \eta_S \bar{n}_S \Delta\nu \quad \text{and} \quad R_B = \bar{n}_B \Delta\nu, \quad (A18b)$$

The measured post-detection visibility $V_{post} := (N_{max,j} - N_{min,j})/(N_{max,j} + N_{min,j})$ may be estimated from each local maximum-minimum pair $j$, with averaging over all $j$. In practice, $V_{post}$ is determined from the fit of the function in eqn. (A17) over all pixels $i$. This would also determine the phase.

However, to bring out the essence of the pixel number dependence of the SNR, we simplify here that $\psi$ is constant over the used number of pixels, and furthermore, that half of the pixels, $m/2$, receive almost full constructive, and the other half almost full destructive interference, independent of the visibility $V$. To consider all pixels for the SNR and not just the maxima and minima, this is achieved by replacing the cosine-modulation by a rectangular modulation with the average amplitude, so that from eq. (A17) it is $\eta_{S,max} = (2/m)g_{max}$ and $\eta_{S,min} = (2/m)g_{min}$, where $g_{max} = (1 + \frac{2}{\pi}V_{pre})/2$ and $g_{min} = (1 - \frac{2}{\pi}V_{pre})/2$, while $\eta_B = 1/m$. For $V_{pre} = 1$ it is $g_{max} \approx 0.818$ and $g_{min} \approx 0.182$. Furthermore, we will use $g_{max} + g_{min} = 1$, $g_{max} - g_{min} = \frac{2}{\pi}V_{pre}$ and $g_{min} \cdot g_{max} = (1 - \frac{4}{\pi^2}V_{pre}^2)/4$. The corresponding averaged pixel counts

$$\bar{N}_X = \frac{2}{m}\sum_{i_X} N_{1px,i} \quad (A19)$$

with $X := min$ or $max$, can be written as

$$\bar{N}_X = \frac{2g_X}{m}\eta_Q R_S \Delta t + \frac{1}{m}\eta_Q R_B \Delta t + \bar{N}_0 \quad (A20)$$

The measured post-detection visibility is then renormalized

$$V_{post} \approx \frac{\pi}{2}(\overline{N_{max}} - \overline{N_{min}})/(\overline{N_{max}} + \overline{N_{min}}) \quad (A21)$$

For the further evaluation we define

$$x := \eta_Q R_S \Delta t, \quad y := \eta_Q R_B \Delta t, \quad z_1 := \bar{N}_0 \quad (A22)$$

so that

$$\bar{N}_X = \frac{2g_X}{m}x + \frac{1}{m}y + z_1 \quad (A23a)$$

$$\overline{N_{max}} - \overline{N_{min}} = \frac{4}{\pi m}V_{pre}\, x \quad (A23b)$$

$$\overline{N_{max}} + \overline{N_{min}} = 2\left(\frac{1}{m}x + \frac{1}{m}y + z_1\right) \quad (A23c)$$

$$V_{post} \approx V_{pre} \frac{x}{(x + y + mz_1)} \quad (A24)$$

We define $\bar{N}_0 := \bar{N}_{dark} + \bar{N}_{read}$. Therein, $\bar{N}_{dark}$ are the dark counts per pixel from electrons thermally excited into the conduction band, as discussed above. This should not depend on the integration time, since the thermal redistribution time is very fast against any technical integration time – this contribution can be significantly reduced by cryogenics. $\bar{N}_{read}$ are the read-out noise electrons per pixel. This is also independent of the integration time and can be quasi eliminated by the EMCCD-principle, which so far is only available in silicon for the visible/NIR range.

To determine the post-detection signal-to-noise ratio of the visibility measurement as $SNR_{V,post}^2 = V_{post}^2 / \overline{(\delta V_{post})^2}$ we have to determine

$$\overline{(\delta V)^2} = \left(\frac{\partial V}{\partial \bar{N}_{max}}\right)^2 \overline{(\delta N_{max})^2} + \left(\frac{\partial V}{\partial \bar{N}_{min}}\right)^2 \overline{(\delta N_{min})^2}$$

$$= \pi^2 \frac{\overline{(N_{min})^2}\overline{(\delta N_{max})^2} + \overline{(N_{max})^2}\overline{(\delta N_{min})^2}}{[\overline{N_{max}} + \overline{N_{min}}]^4} \quad (A25)$$

in which

$$\overline{(\delta N_X)^2} = \sum_{i_X}\left(\frac{\partial \overline{(N_X)}}{\partial N_{1px,i}}\right)^2 (\delta N_{1px,i})^2 \quad (A26)$$

$$= \frac{m}{2}\left(\frac{2}{m}\frac{\partial N_{1px,i}}{\partial N_{1px,i}}\right)^2 (\delta N_{1px,i})^2 = \frac{2}{m}(\delta N_{1px,i})^2$$

the latter according to eq. (A19). The electron number variance at a single pixel is in the integration time $\Delta t$, using eqn. (7):

$$\overline{(\delta N_{1px,i})^2} = \left[\overline{(\delta n_{S,i}')^2} + \Delta M_A \overline{(\delta n_B')^2}\right]\Delta\nu\Delta t + \overline{(\delta N_0)^2} \quad (A27)$$

where we added the variance of the Poissonian dark-count statistics per pixel, $\overline{(\delta N_0)^2} = \overline{N_0}$. Using $\overline{\delta n_i'^2} = \eta_i \overline{n}[1 + \eta_i \overline{n}]$, with $\eta_i := \eta_S \eta_Q \eta_{S,i}$ for the signal, and $\eta_i = \eta_Q \eta_{B,i}$ for the background, and therein again $\eta_{S,max} = 2/m \cdot g_{max}$, $\eta_{S,min} = 2/m \cdot g_{min}$, and $\eta_B = 1/m$, and with the photon rates $R_{S,i} = \eta_S \overline{n}_S \Delta\nu$ and $R_B = \overline{n}_B \Delta\nu$ we get

$$\overline{(\delta N_{1px,i})^2} = \eta_Q[\eta_{S,i} R_S(1 + \eta_S \eta_Q \eta_{S,i}\overline{n}_S) + \eta_B R_B(1 + \eta_Q \eta_B \overline{n}_B)]\Delta t + \overline{N_0} \quad (A28)$$

so that it is

$$\overline{(\delta N_X)^2} = \frac{2}{m}\left[\frac{2\eta_Q}{m} g_X R_S\left(1 + \frac{2\eta_Q \eta_S g_X}{m}\overline{n}_S\right) \cdot \Delta t + \frac{\eta_Q}{m} R_B\left(1 + \frac{\eta_Q}{m}\overline{n}_B\right) \cdot \Delta t + \overline{N_0}\right] \quad (A29)$$

After we applied the approximations

1.) Concentrating this discussion to wavelengths shorter than the far-infrared it is roughly $\overline{n}_B < 5$, and with $\eta_Q \approx 0.7$ it is $(1 + \frac{\eta_Q}{m}\overline{n}_B) < 1.1$ for $m \gg 1$ (e.g. $m \approx 30$ px in the AMBER 3-telescope beam-combiner and 72 px in the 4-telescope instrument MATISSE).

2.) With assuming $\eta_S \ll 1$ in interferometry and again $m \approx \geq 30$, it is clearly also $2\eta_Q\eta_S\overline{n}_S/m \ll 1$. Then we obtain

$$\overline{(\delta N_X)^2} \approx \frac{2}{m}\left[\frac{2\eta_Q}{m} g_X R_S \Delta t + \frac{\eta_Q}{m} R_B \Delta t + \overline{N_0}\right] = \frac{2}{m} \quad (A30)$$

The post-detection SNR of the visibility measurement is therefore combining eqns. (A21) and (A25):

$$SNR_{V,post}^2 = \left(\frac{\pi}{2}\right)^2 \frac{m(\overline{N_{max}} - \overline{N_{min}})^2(\overline{N_{max}} + \overline{N_{min}})}{2\pi^2 \overline{N_{min}} \cdot \overline{N_{max}}} \quad (A31)$$

$$= \frac{(4/\pi^2) V_{pre}^2 x^2(x + y + mz_1)}{(2g_{min}x + y + mz_1)(2g_{max}x + y + mz_1)}$$

$$= \frac{(4/\pi^2) V_{pre}^2 x^2(x + y + mz_1)}{4\left(1 - \frac{4}{\pi^2}V^2\right)x^2 + 2x(y + mz_1) + (y + mz_1)^2} \quad (A32)$$

If all intensity would be on a single pixel, the SNR is

$$(SNR_1)^2 = \overline{(N_{1px})^2}/\overline{(\delta N_{1px})^2} \quad (A33)$$

$$= \frac{(\eta_Q R_S \Delta t)^2}{\eta_Q[R_S(1 + \eta_S \eta_Q \overline{n}_S) + R_B(1 + \eta_Q \overline{n}_B)]\Delta t + \overline{N_0}}$$

with the assumption $\eta_S \eta_Q \overline{n}_S \ll 1$, this gives

$$SNR_1^2 \approx \frac{x^2}{x + y + z_1} \quad (A34)$$

The SNR-loss is therefore:

$$\left(\frac{SNR_{m,post}}{SNR_{1,post}}\right)^2 = \frac{(8/\pi) V_{pre}^2 (x + y + mz_1)(x + y + z_1)}{4\left(1 - \frac{4}{\pi^2}V_{pre}^2\right)x^2 + 2x(y + mz_1) + (y + mz_1)^2} \quad (A35)$$

$$= \frac{V_{pre}^2\left(1 + m\frac{z_1}{x+y}\right)\left(1 + \frac{z_1}{x+y}\right)}{\frac{\left(3 - \frac{16}{\pi^2}V_{pre}^2\right)}{\left(1 + \frac{y}{x}\right)^2} + \left(1 + m\frac{z_1}{x+y}\right)^2} \approx V_{pre}^2 \frac{\left(1 + \frac{z_1}{x+y}\right)}{\left(1 + m\frac{z_1}{x+y}\right)}$$

For $x \ll y$, and with $N_{rad} = x + y < z_1$, $m \gg 1$ it is

$$\frac{SNR_{m,post}}{SNR_{1,post}} \approx V_{pre}\sqrt{\frac{\overline{N}_{rad} + \overline{N}_{read}}{\overline{N}_{rad} + m\overline{N}_{read}}} \quad (A36)$$

giving only an impact when $N_{rad} < N_{read}$.

**8.5. Coherence time of the atmosphere and seeing**

to the inhomogeneous refraction index structure of the atmosphere, the plane wave fronts from space are deformed before reaching a telescope. The Fried-parameter, $r_0$, is the typical transverse length-scale over which the phase of the astronomical light wave front changes by about $\pi$. Interestingly, it was originally introduced by Fried for discussing single-telescope HD [59].

**Single telescopes:**

In case of $r_0 > D$ a single focal spot jitters around its nominal position for which a tip-tilt correction is sufficient. At $r_0 < D$ multiple focal spots move and jump around (speckles) and adaptive optics (AO) is necessary to recover any overlap with the fundamental mode to which a heterodyne receiver is sensitive. In HD, already small phase-front deviations corrupt coupling efficiency (see Ruze-formula [62], derived for radio-telescope dishes), whereas in DD this "just" leads to an enlargement of the focal spot of the point-spread function (PSF) over integration times larger than the coherence time of the atmosphere, and therefore to a reduction of spatial resolution. However, in interferometry this reduced spatial resolution directly leads to a loss of fringe-contrast, because this is equivalent to multiple modes being present. Therefore, fundamental mode filtering is needed, realized with the incorporation of single-mode fibers [38], and so the coherent part of the photons detected in both telescopes suffers the same coupling loss as in heterodyne. Thus, also here AO improves a lot the so-called "throughput" to the beam combiner instrument [60][61] (and offers the opportunity to possibly omit the single-mode fiber filters). The expected drop in coupling efficiency should consequently be the same for both detection schemes, and an AO enhancement for HD should not be considered as an argument against it.

A worst-case estimation of the coupling losses to the fundamental mode may be obtained from the Ruze-formula known from radio astronomy, given by [62]

$$\eta_A = e^{-(4\pi\varepsilon/\lambda)^2} = e^{-4\overline{\Delta\varphi^2}} \quad (A38)$$

where $\varepsilon$ is the standard deviation from the perfect mirror surface (tolerance), or $\overline{\Delta\varphi^2}$ the variance of atmospheric phase variations equivalent to them through $\Delta\varphi = 2\pi\varepsilon/\lambda$. Strictly seen, this formula should be applicable only for high spatial frequencies of perturbations, so for very large telescopes under bad seeing (statistical limit, including high orders in Zernike polynomial), but we assume that after subtraction of the AO corrections, the residual phases are high-frequent enough also for smaller telescopes (except for tip-tilt, but increasingly true for higher-order Zernike polynomials).

In the Kolmogorov-model, $r_0$ scales with wavelength according to

$$r_0 = \left(0.423(2\pi)^2 \sec(\Theta) \int_0^\infty C_n^2(h)\, dh\right)^{-3/5} \cdot \lambda^{6/5} \quad (A39)$$

and has a value of 20cm at a seeing of 0.6 arcsec at 0.5 μm wavelength and zenith ($\Theta = 0$) [63]. The spatial phase spectrum of Kolmogorov-turbulence is given by

$$\Phi(k) = 0.023 r_0^{-5/3} k^{-11/3} \quad (A40)$$

Developed into Zernike polynomial coefficients, $\Phi(R\varrho,\Theta) = \sum_j a_j Z_j(\varrho,\Theta)$, with $\varrho = r/R$, this is

$$\langle a_j^* a_{j\prime}\rangle = \frac{0.046}{\pi}\left(\frac{R}{r_0}\right)^{5/3}[(n+1)(n'+1)]^{1/2}(-1)^{(n'-n)/2}\delta_{mm'}$$
$$\times \int dk\, k^{-8/3}\frac{J_{n+1}(2\pi k)J_{n'+1}(2\pi k)}{k^2} \quad (A41)$$

according to Noll [64], using his numbering. The residual phase errors, $\Delta_J = \langle\Phi^2\rangle - \sum_{j=1}^J \langle|a_j|^2\rangle$, after various degrees of AO-correction, are given by $\Delta_J = \alpha_J (D/r_0)^{5/3}$ [64], with $\alpha_3 = 0.134$ (all up to radial degree $n = 1$, tip-tilt), $\alpha_6 = 0.0648$ (all up to radial degree $n = 2$, defocus and astigmatism), $\alpha_{10} = 0.0401$ (all up to radial degree $n = 3$, coma), $\alpha_{21} = 0.0208$ (all up to radial degree $n = 5$, realized with NAOMI [60]), etc. For the discussed worst-case estimation, we insert this now into eqn. (A38) which we plot in Fig. 8:

$$\eta_A(\lambda)/\eta_0(\lambda) \approx e^{-4\Delta_J} = e^{-4\alpha_J(D/r_0)^{5/3}} \quad (A42)$$
$$= exp\left[-4\alpha_J \left(D/r_{0,0.5\mu m}\right)^{5/3}(0.5\mu m/\lambda)^2\right]$$

with $r_{0,\lambda} = r_{0,0.5\mu m} \cdot (\lambda/0.5\mu m)^{6/5}$ and where we assume the maximum possible coupling to a single-mode fiber or a Gaussian receiver mode being $\eta_0(\lambda) \approx 0.7$ (to be optimized at each wavelength) in a Cassegrain design with a focal ratio of 6 and a typical central blockage of $\alpha = 0.2$ [65]. We can see that for AT-sized telescopes tip-tilt correction should be satisfactory in the mid-infrared range ($\lambda > 5$ μm, whereas for UT-sized ones AO is needed.

**Interferometry:**
The residual after a full AO-correction is the differential piston movements of the phase between the telescopes. Without further measures, in interferometry it is necessary to keep the integration time shorter than the atmospheric coherence time, to not lose the interference fringes.
In a frozen-atmosphere approximation of a single turbulence layer the spatial structure of the phase pattern is assumed to move over the telescope or the interferometer array in a static manner with the wind-speed, and therefore the turbulence coherence time is $\tau_{coh} = 0.314\, r_0/v$ [10]. At the high wind speeds $v$ usually seen at the VLTI-site, it can be shorter than 10 ms in the K-band ($\lambda = 2.2\ \mu m$).

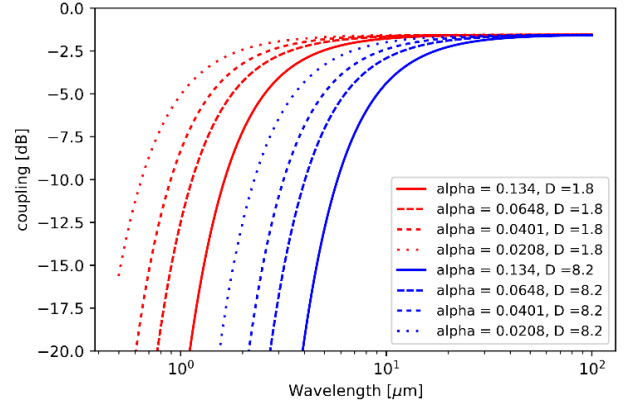

*Fig. 8: Ruze-type coupling efficiency over wavelength with various degrees of wavefront correction for the Paranal telescopes.*

In DD interferometry, fringe trackers can be employed to achieve longer integration times than the atmospheric coherence time by looking at stronger point sources nearby the science target. In HD interferometry, however, we can also use the triple-correlation product ("bispectrum") with its closure-phase as a primary quantity as argued in the following, while it is a secondary quantity in DD, because it can be determined from the fringe patterns only after integration on the detectors, wherein the closure-phase can be determined only from the intensities and displacements of the fringes.

This is so interesting because the closure-phase-property applies in any moment so that in HD the correlator can determine the complex triple-product already after any FFT-time interval, which can be on the microseconds-scale, and could then integrate this quantity. Due to the constantly applying closure-phase property, this integration, executed over all available baseline-triangles, can in principle retrieve the individual baseline-visibilities after algebraic operations (in case of more than three telescopes) without SNR-loss during longer integrations than the atmospheric coherence time. The original term $\sqrt{\Delta\nu\Delta t}$ should then stay valid also for the determination of the individual visibilities even without stabilizing the phases on the individual baselines, which would constitute an important advantage of HD over DD. In DD this is not possible, because the time-integration on the CCD (delivering the two-telescope visibilities and phases, but with a visibility-loss, see eq. A36 and A45) occurs before the calculation of the closure-phase. It was already demonstrated in heterodyne interferometry that closure phase measurements (after time-integration) are beneficial for image reconstruction by adding constraints to the two-telescope correlation products (visibilities and phases) [66].

Townes et al. [28], [29] argued that for $\Delta t > \tau_{coh}$ in cross-correlation the atmospheric coherence time influences the dependence on the integration time in the way of the replacement $\Delta t \to \sqrt{\Delta t\, \tau_{coh}}$, so that we have

$$\begin{array}{ll} SNR_{CC} \propto \sqrt{\Delta\nu\, \Delta t} & \text{for } \Delta t \leq \tau_{coh} \\ SNR_{CC} \propto \sqrt{\Delta\nu}\, \sqrt[4]{\Delta t\, \tau_{coh}} & \text{for } \Delta t > \tau_{coh} \end{array} \quad (A43)$$

This allows larger integration times than $\tau_{coh}$ which are, however, less effective due to the slower time-dependence. Please also note the remarks on the CC-SNR in appendix 8.3 and in section 2.

For short integration times around the atmospheric coherence time, we can argue with the following, in the way previously done in [67]. From the signal and noise phasor diagram it can be inferred that the phase uncertainty in a single spectral channel is given by

$$\delta\varphi_{1ch}^2 = \frac{1}{SNR_{1ch}} \quad (A44)$$

using our squared SNR pre-detection definition. A phase detection is possible at all, and therefore a visibility measurement, only if $\delta\varphi < 1$. If this condition is given for each spectral channel, then the total phase precision can be improved by averaging the phases from all channels, and the result has then the uncertainty $\delta\varphi_{Nch}^2 = \delta\varphi_{1ch}^2/N_{ch}$, see [67]. Due to the phase fluctuations of width $\delta\varphi$ around $\varphi_0$ during the integration time, the expected integrated interference signal amplitude diminishes according to

$$S(t) = \int_{-\infty}^{\infty} \frac{V}{\delta\varphi\sqrt{\pi}} e^{-\left(\frac{\varphi-\varphi_0}{\delta\varphi}\right)^2} \cos(\Omega t + \varphi)\, d\varphi$$
$$= V e^{-\delta\varphi^2} \cos(\Omega t + \varphi_0) \quad (A45)$$

with $V$ the visibility, where the coordinate $t$ is either the time in HD, or the spatial pixel coordinate in DD. If we assume a random walk of the phase in time, which is the reason for the Gaussian form assumed in eq. (A45), then it is $\delta\varphi^2 = \Delta t/\tau_{coh}$, where we define the atmospheric coherence time with $\delta\varphi = 1$ (e.g. in the K-band it is around 10 ms), and we obtain a corrected SNR as

$$SNR_{CC,\varphi} \coloneqq SNR_{CC}\, e^{-\Delta t/\tau_{coh}} = SNR_{CC}\, e^{-1/SNR_{CC}} \quad (A46)$$

This produces a slight collapse of the fringe-SNR to 0 dB where it would be still 2-3 dB without this effect. A fringe-tracker in the DD scheme is supposed to avoid this. For HD it would be useful to confirm experimentally whether the above-proposed bi-spectrum closure-phase integration scheme would bypass this collapse and would allow to integrate longer even without a fringe-tracker.